%Paper: hep-th/9510130
%From: Neil Lambert <N.Lambert@damtp.cam.ac.uk>
%Date: Wed, 18 Oct 95 12:38:35 BST
%Date (revised): Wed, 18 Oct 95 15:58:18 BST

\input phyzzx
\input epsf
%\epsfverbosetrue

\twelvepoint

\font\bb=msbm10 scaled 1200

\def\dminus{{\partial}_=}
\def\dmm{{\partial}_=}
\def\dpp{{\partial}_{\ne}}

\def\delmm{\nabla^{(+)}_=}
\def\delpp{\nabla^{(-)}_+}

\def\Dph{\hat{D}_+}
\def\Dp{D_+}
\def\Psima{\Psi_{-}^{a}}
\def\Psimb{\Psi_{-}^{b}}

\def\thetap{\theta^{+}}
\def\bG{\bar \Gamma}
\def\bg{\bar g}
\def\bb{\bar b}
\def\bA{\bar A}
\def\bs{\bar s}
\def\a{\alpha'}
\def\cX{\cal X}

\def\ssm{ supersymmetric sigma model}
\def\dalemb#1#2{{\vbox{\hrule height .#2pt
        \hbox{\vrule width.#2pt height#1pt \kern#1pt
                \vrule width.#2pt}
        \hrule height.#2pt}}}
\def\square{\mathord{\dalemb{7.9}{10}\hbox{\hskip1pt}}}

\REF\Alvarezo{L. Alvarez-Gaum{\'e}, D.Z. Freedman and S. Mukhi, Ann. Phys.
{\bf 134} (1983) 85}
\REF\Hullo{C.M. Hull and E. Witten, Phys. Lett. {\bf 60B} (1985) 398}
\REF\Hullt{C.M. Hull and P.K. Townsend, Phys. Lett. {\bf 178B} (1986) 187}
\REF\Gates{S.J. Gates, JR., M.T. Grisaru, L. Mezincescu and P.K. Townsend,
Nucl. Phys. {\bf B286} (1987) 1}
\REF\Hullth{C.M. Hull, G. Papadopoulos and P.K. Townsend, Phys. Lett. {\bf
316B}
(1993) 291}
\REF\GP{G. Papadopoulos and P.K. Townsend, Class. Quan. Grav. {\bf 11} (1994)
515}
\REF\Witten{E. Witten, Nucl. Phys. {\bf B403} (1993) 159}
\REF\Wittentwo{E. Witten, J. Geom. Phys. {\bf 15 } (1995) 215}
\REF\Alvarezt{L. Alvarez-Gaum{\'e}
and D.Z. Freedman, Comm. Math. Phys {\bf 91} (1983) 87}
\REF\Pap{G. Papadopoulos and P.K. Townsend, Nucl. Phys. {\bf B444} (1995) 245 }
\REF\NDL{N.D. Lambert, "Quantizing the (0,4) Supersymmetric ADHM Sigma Model",
DAMTP-R/95/43, hep-th/9508039}
\REF\Pault{C.M. Hull and P.K. Townsend, Phys. Lett. {\bf
191B} (1987) 115}
\REF\All{R.W. Allen and D.R.T. Jones, Nucl. Phys. {\bf B303} (1988) 271}
\REF\Jones{D.R.T. Jones, Phys. Lett. {\bf 192B} (1987) 391}
\REF\VdV{B.E. Fridling and A.E.M. van de Ven, Nucl. Phys. {\bf B268} (1986)
719}
\REF\GM{E. Guadagnini and M. Mintchev, Phys. Lett.  {\bf 186B} (1987) 173}
\REF\Foake{A.P. Foakes, N. Mohammedi and D.A. Ross, Phys. Lett. {\bf 206B}
(1988) 335}
\REF\Foakes{A.P. Foakes, N. Mohammedi and D.A. Ross, Nucl. Phys. {\bf B310}
(1988) 335}
\REF\Ell{U. Ellwanger, J. Fuchs and M.G. Schmidt, Phys. Lett. {\bf 203B} (1988)
244}
\REF\Howet{P.S. Howe and G. Papadopoulos, Nucl. Phys. {\bf B289} (1987) 264}
\REF\Stelle{P.S. Howe, G. Papadopoulos and K.S. Stelle, Nucl. Phys. {\bf B296}
(1988) 26}
\REF\Grisaru{M.T. Grisaru, M. Ro$\check c$ek and W. Siegel, Nucl. Phys {\bf
B159}
(1979) 429}
\REF\HW{P.Howe and P. West, Phys. Lett. {\bf 156B} (1985) 335}
\REF\Mukhio{S. Mukhi, Nucl. Phys. {\bf B264} (1986) 640}
\REF\Howeo{P.S. Howe and G. Papadopoulos, Class. Quantum Grav. {\bf 4} (1987)
1749}
\REF\Callan{C.G. Callan, D. Friedan, E.J. Martinec and M.J.
Perry, Nucl. Phys. {\bf B262} (1985) 593}
\REF\Paffuti{G. Curci and G. Paffuti, Nucl. Phys. {\bf B286} (1987) 399}
\REF\Osborn{H. Osborn, Ann. Phys. {\bf B200} (1990) 1}
\REF\Ross{D.A Ross, Nucl. Phys. {\bf B286} (1987) 93}
\REF\Hamada{K. Hamada, J. Kodaira and J. Saito, Nucl. Phys. {\bf B297} (1988)
637}
\REF\Tseytlin{A.A. Tseytlin, Nucl. Phys. {\bf B294} (1987) 383}
\REF\Jack{I. Jack, D.R.T. Jones and D.A. Ross, Nucl. Phys. {\bf B307} (1988)
531}
\REF\Papa{G. Papadopoulos, "Global Aspects of Sigma Models with
Torsion", hep-th/940617}
\REF\Mur{T. Murphy and L. O'Raifeartaigh, Nucl. Phys. {\bf B218} (1981) 484}
\REF\Sal{A.C. Davis, P. Salomonson and J.W. van Holten, Phys. Lett. {\bf 113B}
(1982)
472}
\REF\GVZ{M. Grisaru, Nucl. Phys. {\bf B277} (1986) 409}
\REF\Hullf{C.M. Hull, Phys. Lett. {\bf 167B} (1986) 51}
\REF\Howthree{P.S. Howe and G. Papadopoulos, Nucl. Phys. {\bf B381} (1992) 360}
\REF\Muhkit{S. Mukhi, Phys. Lett. {\bf 162B} (1985) 345}

\Pubnum={DAMTP R/94/42\cr hep-th/9510130}

%\date

\titlepage

\title{\bf Two Loop Renormalization of Massive (p,q) Supersymmetric Sigma
Models}

\centerline{N.D. Lambert\foot{nl10000@amtp.cam.ac.uk}}

\address{D.A.M.T.P., Silver Street\break
         University of Cambridge\break
         Cambridge, CB3 9EW\break
         England}

\vfil

\abstract

We calculate the $\beta$-functions of the general massive (p,q)
supersymmetric sigma model to two loop order using (1,0) superfields. The
conditions for finiteness are discussed in relation to
(p,q) supersymmetry. We also calculate the effective potential using
component fields to one loop order and consider the possibility of perturbative
breaking of supersymmetry. The effect of one loop finite local counter terms
and the
ultra-violet behaviour of the off-shell (p,q) models to all orders in
perturbation
theory are also addressed.

\endpage

\chapter{Introduction}

In the past massless supersymmetric sigma models have been extensively studied
in
connection with superstring theory (see [\Alvarezo,\Hullo,\Hullt,\Gates] and
the
references therein) and also for their relation to differential geometry and
topology. More recently there has been interest in models which include a
potential
for the bosonic fields (so called massive sigma models)
[\Hullth,\GP,\Witten,\Wittentwo]. Although originally studied in connection
with the
need to eliminate the infrared divergences occurring in the massless models
[\Alvarezo,\Alvarezt], the existence of a potential for the bosonic fields
allows for
interesting nonperturbative effects, such as soliton solutions interpolating
between
the different vacua of the theory [\Pap]. The phenomena which occur in these
models
are analogous to those appearing in 4 dimensional Yang-Mills-Higgs theories. In
addition, certain classes of these theories have interesting renormalization
properties, which have attracted attention due to their relationship to (2,0)
Landau-Ginzburg models and the moduli space of superstring vacua. In particular
relatively simply massive linear sigma models can flow under the
renormalization
group to highly non trivial conformally invariant field theories in the
infrared
limit [\Wittentwo,\NDL].

In the literature there exist many two loop calculations
of generalized sigma models discussing several issues. A brief but
certainly not complete listing is as follows. Reference [\Alvarezo]
calculated the two loop $\beta$-functions of the purely metric bosonic and
(1,1)
supersymmetric sigma model. The $\beta$-function for bosonic model with with
torsion
was calculated in [\Pault]. These were followed by other calculations in the
bosonic
and (1,1) supersymmetric models by several authors (including discussions of
the
ambiguities associated with dimensional regularization). In particular
[\All,\Jones]
used component fields while [\VdV,\GM] used (1,1) superfields. The (1,0)
supersymmetric sigma model $\beta$-functions have been calculated to three
loops for
a vanishing antisymmetric tensor [\Foake,\Foakes] using component fields, and
for a
purely gauge field background by [\Ell] with (1,0) superfields. To the best of
our
knowledge, there exists no two loop calculation using (1,0) superfields where
all the
background fields are non vanishing.

In this paper we study, to two loop order, the renormalization and finiteness
of the
general off-shell (p,q) supersymmetric massive sigma model, on flat two
dimensional
Minkowski Space. We will be particularly interested in the addition of  mass
terms,
not considered in previous supersymmetric calculations. We use (1,0) superspace
perturbation theory and no assumptions are made about any of the
background fields. It should be noted that even in the case of
(1,1) supersymmetry the most general form of the potential, where a central
charge
appears in the supersymmetry algebra, cannot be expressed with (standard) (1,1)
superfields, but can with (1,0) superfields [\Hullth]. Our work therefore
applies to
the most general masssive (p,q) supersymmetric sigma model.

After introducing the model in the next section we discuss the background field
quantization method in section 3. Then, in section 4, this method is used to
calculate the one loop contributions to the $\beta$-functions and the effective
potential. We discuss the conditions for finiteness in  relation to (2,0) and
(1,1)
supersymmetry and the possibility of the perturbative breaking of
supersymmetry. In
section 5 the $\beta$-functions are calculated to two loop order and the effect
of
one loop finite local counter terms discussed. Finally, in section 6 we discuss
the ultra-violet behaviour of
the general (p,q) supersymmetric model to all orders of perturbation theory.
Here we
show that there are no mass renormalizations to any order of perturbation
theory for
the (2,0) supersymmetric models. We conclude with a brief discussion of the
conditions for conformal invariance and the relevance of massive sigma models
to
string theory.

\chapter{The (1,0) Supersymmetric Massive Sigma Model}

The massive {\ssm} is defined by maps from flat (1,0) superspace
$\Sigma^{(1,0)}$
into a rank n vector bundle $\,\Xi\,$ over a d dimensional Riemannian target
manifold
{\cal M}. The field content consists of d even scalar superfields
$\Phi^i(x,\thetap){\ }(i=1,...,d)$ mapping $\Sigma^{(1,0)}$ into {\cal M} and
and n
odd right handed spinor superfields $\Psi^a_-(x,\thetap){\ }(a=1,...,n)$ which
map
$\Sigma^{(1,0)}$ into the pull back of $\,\Xi\,$ by $\Phi$ (ie. they map into
the
fibre above a point $\Phi^i(x,\thetap)$ in {\cal M}). The $\Phi^i$ can be
thought of
as the coordinates on the manifold {\cal M} while $\Psima$ is a section of $S_-
\otimes \Phi^*\Xi\ $, where $S_-$ is the (right handed) spin bundle over
$\Sigma^{(1,0)}$.

In order to define the (1,0) supersymmetry algebra we introduce light cone
coordinates $(x_{\ne} ,x_=)$ defined as
$$
x_{\ne} = {1\over{\sqrt{2}}}(x^0+x^1)\ ,\ \ \ \ \ \ \
x_= = {1\over{\sqrt{2}}}(x^0-x^1)\ ;
$$
$$
\dpp = {1\over{\sqrt{2}}}(\partial_0+\partial_1)\ ,\ \ \ \ \ \ \
\dminus = {1\over{\sqrt{2}}}(\partial_0-\partial_1) \ .
$$
Hence $\square=\partial^{\mu}\partial_{\mu}=2\dpp\dminus$. The "stacked"
subscripts
count the Lorentz weight of a variable (ie. their transformation properties
under the
Lorentz group SO(1,1)) as in [\Howet]. Vector components have Lorentz weight
$\pm2$
while those of spinors have Lorentz weight $\pm1$. The (1,0) superspace
covariant
derivative $\Dp$ is defined as,
$$
\Dp  = {{\partial} \over{\partial{\thetap}}}
+ i\thetap \dpp\ ,
\eqn\Dplus
$$
so that $D_+^2 = i\dpp$. Throughout this paper a hat over a derivative refers
to the
covariant derivative  induced by an $\Xi$ connection $A^a_{i{\ }b}$, on $\cal
M\ $.
For example $$
\Dph\Psima=\Dp\Psima + \Dp\Phi^iA^a_{i{\ }b}\Psimb\ , \eqn\Dhat
$$
where here we have pulled back the covariant derivative to $\Sigma^{(1,0)}$.

Given these definitions the action for the massive{\ssm} is written as
$$
S={-i}\int\! d^2xd\thetap\left\{ (g_{ij}+b_{ij})\Dp\Phi^i\dminus\Phi^j
+ i h_{ab}\Psima\Dph\Psimb + ims_a \Psima\right\}\ ,
\eqn\action
$$
where $g_{ij}$ and $b_{ij}$ are metric and antisymmetric tensor fields on {\cal
M}
respectively. $h_{ab}$ is a fibre metric which, following [\Hullth,\GP], we can
assume
without loss of generality to be covariantly constant with respect to the fibre
covariant derivative
$\hat{\nabla}_i$. The parameter m is a constant of mass dimension one and $s^a$
is an arbitrary section of $\,\Xi\,$. We denote by $S_\Phi$, $S_\Psi$ and $S_m$
the
three terms in {\action} respectively. Throughout most this paper the fibre
metric assumed implicitly. Repeated vector bundle indices will still be summed
over
using $h_{ab}$.

When expanded in terms of the component fields, $\Phi^i=\phi^i+\thetap\eta^i_+$
and
$\Psima=\psi^a_-+\thetap F^a$, the action {\action} can be reduced to an
integral over
two dimensional Minkowski space by integrating over $\thetap$. After
eliminating the
auxiliary fields $F^a$ by their equations of motion, one recovers the standard
supersymmetric sigma model but with a potential for the bosonic fields
[\Hullth,\GP]
$$\eqalign{
S= \int\! d^2x & \left\{ (g_{ij}+b_{ij})\dpp\phi^i\dminus\phi^j
+ig_{ij}\eta^i_+\nabla^{(+)}_=\eta^j_+ -
ih_{ab}\psi^a_-\hat{\nabla}_{\neq}\psi^b_-
\right. \cr & \left.
-{1\over2}\psi^a_-\psi^b_- F_{abij}\eta^i_+\eta^j_+ +
m\hat{\nabla}_is_a\eta^i_+\psi^a_- -{1\over4}m^2h^{ab}s_as_b \right\} \ , \cr
}\eqn\potential
$$
where the covariant derivatives in \potential\  are defined in section 3
below.
Furthermore, supersymmetry requires that $h_{ab}$ is positive definite so that
the
potential $V(\phi)={1\over4}h^{ab}s_as_b$, and hence the energy, is positive. A
generic section will have isolated zeros. The constant values of the scalar
fields at
these points constitute the classical vacuum configurations. In the case of
(1,1)
supersymmetry at least one of these vacua must survive as the quantum zero
energy vacuum unless the Witten index vanishes, which is possible only for
target
space with vanishing Euler number. In these special cases, which include models
with
a group manifold for the target space, the the section $s_a$ may have several
or no
zeros, depending on the choice of parameters defining the potential [\Pap].

\chapter{Background Field Quantization}

In order to perform the quantization and renormalization of the (1,0) {\ssm} it
is
most convenient to use background field method [\Stelle] together with (1,0)
superspace perturbation theory [\Hullt,\Gates,\Ell,\Grisaru,\HW]. (1,0)
superfield
methods allows for the general (p,q) massive supersymmetric model to be
considered and
therefore is the most useful for our purposes. All previous calculations of
sigma
model $\beta$-functions have only used either component field or, in the case
of (1,1)
supersymmetry, (1,1) superfields (with the exception of [\Ell] which also uses
(1,0)
superfields but only includes background gauge fields).

The background field method
entails splitting up the fields into "background" and "quantum" parts and
integrating
over the quantum fields in the generating functional. This is achieved by
summing
over all graphs with no external quantum legs. To renormalize the theory to two
loop
order the action is expanded to fourth order in the quantum fields so that the
relevant vertices can be determined. By far the simplest way to expand the
action is
to use the algorithm introduced by Mukhi [\Mukhio],
exploiting the manifest tensorial structure of the action. Here we will
generalize
this algorithm so as to include the torsion implicitly into the terms in the
expansion which greatly simplifies the calculation.

First, we wish to spilt up the fields $\Phi^i_{total}$ and ${\Psima}_{total}$,
which appear in the action {\action}, into background and quantum fields. In
order
to construct a manifestly covariant expansion of the action, the quantum fields
must transform covariantly under coordinate and gauge transformations.
If we naively expand the action {\action} around fixed background fields
$\Phi^i$ and
$\Psima$ to
$$
S(\Phi^i_{total},{\Psima}_{total})= S(\Phi^i+\pi^i,\Psima+\chi^a_-)
$$
we immediately see that the fields $\pi^i$ and $\chi^a_-$ are not vectors under
coordinate transformations. Thus, in order to perform an expansion which is
manifestly coordinate invariant, we must find another definition for the
quantum
fields. There is a well known procedure for doing this [\Alvarezo]. For any
background field $\Phi^i$ and total field $\Phi^i_{total}$, sufficiently close
to it,
there exists a unique geodesic of unit length, passing through each point. We
define the quantum field $\xi^i$ as the tangent to this geodesic at the point
$\Phi^i$. Specifically, if $\Phi^i(s)$ is the geodesic with $\Phi^i(0)=\Phi^i$,
the
background field, and $\Phi^i(1)=\Phi^i_{total}$ the total field appearing in
the
action. We then have
$$
{d^2\Phi^i(s)\over{d^2s}}+\Gamma^i_{jk}{d\Phi^j(s)\over{ds}}{d\Phi^k(s)\over{ds}}=0\
{}.
$$ Then we define the quantum field as $\xi^i =
{d\Phi^i(s)\over{ds}}\vert_{s=0}\
$. Similarly, $\zeta^a_-$ is the tangent to the geodesic $\Psima(s)$, joining
$\Psima$ to the total field appearing in the action ${\Psima}_{total}$. The
geodesic
equation for $\Psima(s)$ is
$$
{d^2\Psima(s)\over{d^2s}}+A^a_{i\ b}{\Psimb(s)}{d\Phi^i(s)\over{ds}}=0\ ,
$$
and the quantum field is defined to be $\zeta^a_- =
{d\Psima(s)\over{ds}}{\vert}_{s=0}\
$. This non trivial definition will in general lead to a non linear
renormalization of
the quantum fields at higher loops [\Stelle] but this will not affect the
calculation
here.

The advantage of using the background field method is that, due to a shift
symmetry
in the choice of the  background/quantum split, the symmetries of the action
are
preserved under quantization [\Stelle]. The action \action\
is invariant under general coordinate transformations of the target space $\cal
M$
and gauge transformations of the vector bundle $\Xi$. Although it should be
noted
that these symmetries do not give rise to associated Noether charges as the
spacetime fields $g_{ij},\ b_{ij},\ A_{i\ b}^a,\ {\rm and\ }s_a$ (ie. coupling
constants) must be varied in addition to the worldsheet fields $\Phi^i {\rm \
and\ }
\Psima$. These symmetries are also subject to anomalies in the quantum theory
which
can be removed as we will discuss below. By using the background field method
this
tensorial structure of the action can be maintained in perturbation theory.
This
allows us to use the methods of standard tensor analysis and choose to work in
a
frame where the expressions are relatively simple, such as normal coordinates,
and
then transform back to a general frame for the final answer. In normal
coordinates,
one finds  $\xi^i=\pi^i$ and $\zeta^a_-=\chi^a_-$ [\Alvarezo], and the
expansion can
be readily performed.

While this method is certainly sufficient, it requires rather lengthy
calculations.
To avoid such tedious algebra we use a further simplification which allows one
to
write down the (n+1)th term in the expansion, in covariant form, directly from
the
nth term. This method utilizes the algorithm first worked out by Mukhi
[\Mukhio]. In
Mukhi's  algorithm the various geometric objects (tensors and covariant
derivatives)
appearing  in the nth term of the expansion combine simply to give the (n+1)th
term
without resorting to the use of normal coordinates.

It is useful at this point to introduce some additional geometric structures on
the
target space bundle. First, as the quantum fields $\xi^i$ are target manifold
vectors,
we need to replace the derivative operators by the pull back of $\nabla$
on {\cal M} to $\Sigma^{(1,0)}$ when acting on $\xi^i$
$$\eqalign{
\nabla_{\ne}\xi^i&=\partial_{\ne}\xi^i+\Gamma^i_{jk}\partial_{\ne}\Phi^j\xi^k
\cr
\nabla_=\xi^i&=\partial_=\xi^i+\Gamma^i_{jk}\partial_=\Phi^j\xi^k \cr
\nabla_+\xi^i&=D_+\xi^i+\Gamma^i_{jk}D_+\Phi^j\xi^k \ , \cr
}\eqn\trico
$$
where $\Gamma^i_{jk}$ are the connection coefficients of $\nabla$.

Under the transformation $b_{ij} \rightarrow b_{ij}+\partial_{[i}\lambda_{j]}$
where
$\lambda_i$ is an arbitrary 1-form, the action {\action} changes by a surface
term.
Thus there is a gauge freedom in the choice of $b_{ij}$ and hence we define the
(gauge
invariant) field strength as
$$
H_{ijk}={3\over2}\partial_{[i}b_{jk]} \ .
\eqn\H
$$
In the action {\potential} the antisymmetric field strength $H_{ijk}$ acts as
a torsion for the covariant derivative. We
therefore define the connection coefficients and corresponding Riemann tensor
with torsion as
$$\eqalign{
{\Gamma^{(+)}}^i_{jk}&=\Gamma^i_{\ jk} + H^i_{\ jk} \cr
{R^{(+)}}^i_{jkl}&=R^i_{jkl} - \nabla_l H^i_{\ kj} + \nabla_k H^i_{\ lj}
+ H_{mlj}{H^{mi}}_k - H_{mkj}{H^{mi}}_l \ , \cr}
\eqn\plus
$$
and similarly for ${\Gamma^{(-)}}^i_{jk}$ and ${R^{(-)}}^i_{jkl}$ with
$H_{ijk}$
replaced by $-H_{ijk}$. It follows immediately from {\plus} that
$$
R^{(\pm)}_{ij}
= R_{ij} \mp \nabla^kH_{kij}-H_{ikl}H_j^{\ kl}\ , \eqn\Rpij
$$
and hence
$$
R^{(\pm)}_{(ij)}=R_{ij}-H_{ikl}H_j^{\ kl}\ ,\ \ \ \ \ \ \ \ \
R^{(\pm)}_{[ij]}=\mp \nabla^kH_{kij}\ .
\eqn\Revenodd
$$
The standard symmetries of the Riemann tensor (not associated with the
Levi-Civita
connection) are for ${R^{(\pm)}}^i_{jkl}$
$$\eqalign{
R^{(\pm)}_{[ijk]l}&=\mp {2\over3}\nabla^{(\pm)}_{l}H_{ijk} \ , \cr
R^{(\pm)}_{(ij)kl}&=R^{(\pm)}_{ij(kl)}=0 \ . \cr}
\eqn\syms
$$
While the Bianchi identity is
$$
\nabla^{(\pm)}_{[i}R^{(\pm)}_{|mn|jk]}
 = \mp 2 H_{[ij}^{\ \ \ p}R^{(\pm)}_{|mnp|k]} \ .
\eqn\Bianchi
$$
As a result of their antisymmetry, $H_{ijk}$ and $b_{ij}$ compose the
components
of differential forms  $H=H_{ijk}d\Phi^i \wedge d\Phi^j \wedge d\Phi^k$ and
$b=b_{ij}d\Phi^i \wedge d\Phi^j$. Hence it follows from \H\ that
$dH=\nabla_{[i}H_{jkl]}=0$. This identity for $H_{ijk}\ $, along with the above
symmetries of the Riemann tensor $R_{ijkl}$, yield the following useful
relation
$$
R^{(+)}_{ijkl}=R^{(-)}_{klij}\ .
\eqn\iden
$$
It will be convenient to define two different derivative operators with torsion
for
the  left and right moving $\xi^i$ fields. These covariant derivatives are
defined from the derivatives in {\trico} and connections in {\plus} as
$$\eqalign{
\nabla^{(+)}_=\xi^i&=\nabla_=\xi^i+H^{i}_{\ jk}\partial_=\Phi^j\xi^k\ , \cr
\nabla^{(-)}_+\xi^i&=\nabla_+\xi^i-H^{i}_{\ jk}D_+\Phi^j\xi^k\ . \cr
}\eqn\Dtwist
$$

Mukhi's method [\Mukhio] is based on the observation that the nth term ${\cal
L}^{n}$ in the expansion of the action can be calculated from ${\cal L}^{n-1}$
by acting with the operator $\xi^i{\hat {\cal D}}_i^1$,
$$\eqalign{
{\cal L}^{n}(x_2,\thetap_2)&={1\over n} \int\! d^2x_1d\thetap_1\left\{
\xi^i(x_1,\thetap_1){\hat {\cal D}}_i^1{\cal L}^{n-1}(x_2,\thetap_2) \right\}
\cr
& \equiv{1\over n}{\cal A}({\cal L}^{n-1}) \ ,\cr}
$$
where ${\hat {\cal D}}_i^1$ is the covariant functional derivative with respect
to
$\xi^i(x_1,\thetap_1)$, defined on tensors $T_{klm...}^{abc...}$ by
$$
\hat{{\cal D}}_i^1T_{klm...}^{abc...}(x_2,\thetap_2)=\hat{\nabla}_i
T_{klm...}(x_1,\thetap_1) \delta^2(x_1-x_2)\delta(\thetap_1-\thetap_2)\ .
$$

With the geometrical structures defined above, it is not much trouble to prove
the
following generalization of Mukhi's algorithm [\Mukhio] for the sigma model on
(1,0)
superspace with torsion
$$\eqalign{
{\cal A}(\xi^i)&=0 \  \cr
{\cal A}(\zeta_-^a)&=0 \  \cr
{\cal A}(\Phi^i) &= \xi^i\  \cr
{\cal A}(\Psima) &= \zeta^a_-\ \cr
{\cal A}(D_+\Phi^i) &= \nabla^{(-)}_+\xi^i+H^i_{\
jk}D_+\Phi^j\xi^k\  \cr
{\cal A}(\partial_=\Phi^i) &=\nabla^{(+)}_=\xi^i-H^i_{\
jk}\partial_=\Phi^j\xi^k\
\cr {\cal A}({\hat\nabla}_+\Psima)&=
{\hat\nabla}_+\zeta^a_--F^{ab}_{ij}D_+\Phi^i\Psimb\xi^j\ \cr
{\cal A}(\nabla^{(-)}_+\xi^i) &= {R^{(-)}}^i_{(kl)j}D_+\Phi^j\xi^k\xi^l  -
H^i_{\ jk}\nabla^{(-)}_+\xi^j\xi^k  \cr
{\cal A}(\nabla^{(+)}_=\xi^i) &=
{R^{(+)}}^i_{(kl)j}\partial_=\Phi^j\xi^k\xi^l + H^i_{\
jk}\nabla^{(+)}_=\xi^j\xi^k
\cr {\cal A}({\hat\nabla}_+\zeta^a_-) &= -F^{ab}_{ij}D_+\Phi^i\xi^j\zeta^b_-
\cr
{\cal A}(T^{a_1...a_m}_{i_1...i_n}) &= \hat{\nabla}_k T^{a_1...a_m}_{i_1...i_n}
\xi^k\ . \cr } \eqn\rules
$$
In the above rules we have incorporated the torsion implicitly into
the geometrical terms in the algorithm developed by Mukhi [\Mukhio], as applied
to (1,0) superspace. This will greatly simplify the algebra.

\chapter{One Loop Renormalization}

The action {\action}, expanded to second order in the quantum fields using the
algorithm described in the last section is
$$
\eqalign{
S^{(2)}=&{-i}\int\! d^2xd\thetap\{ g_{ij}\nabla^{(-)}_+\xi^i\nabla^{(+)}_=\xi^j
+ i\zeta_-^a{\hat \nabla}_{+}\zeta_-^a + im\hat{\nabla}_is_a\xi^i\zeta_-^a
\cr
&+R^{(+)}_{i(kl)j}\Dp\Phi^i\dminus\Phi^j\xi^k\xi^l
+{1\over2}im\hat{\nabla}_{(i}\hat{\nabla}_{j)}s_a\Psima\xi^i\xi^j
-2i{\Psima}F^{ab}_{ij}\Dp\Phi^i\xi^j\zeta_-^b
\cr
&+{1\over2}i\Psima{\Psimb}F^{ab}_{ij}\delpp\xi^i\xi^j
-{1\over2}i\Psima\Psimb\hat{\nabla}^{(+)}_{(k}F^{ab}_{j)i}\Dp\Phi^i\xi^j\xi^k\}\ .
}
\eqn\actiontwo
$$
The first two terms yield the
propagator for the $\xi^i$ and $\zeta^a$ fields. However, before the
propagators can be
read off, $g_{ij}$ and $h_{ab}$ must be absorbed into a redefinition of the
quantum fields. This is done by referring all the fields to a vielbein frame
$$\eqalign{
g_{ij}&=e^{\ I}_ie^{\ J}_j\delta_{IJ} \ , \cr
h_{ab}&=\hat{e}^{\ A}_a\hat{e}^{\ B}_b\delta_{AB} \ , \cr}
$$
where the fields are then $\xi^i=e^i_{\ I}\xi^I$, $\zeta^a=\hat{e}^a_{\
A}\zeta^A$. We will consider the quantum fields to be in vielbein frames
implicitly and write all tensorial expressions in terms of coordinate frames.

To regulate the ultra-violet divergences we use dimensional regularization and
then
renormalize with modified minimal subtraction. However, the (1,0) supersymmetry
algebra can
not be defined in $2+\epsilon$ dimensions [\Hullo] as there is no clear
distinction between
left and right movers there. To enable us to use dimensional regularization we
follow the
methods introduced in [\Gates] whereby we perform the $D_+$ algebra first and
integrate over all but one of the Grasmannian coordinates, in order to obtain
integrals which can be written in a manifestly Lorentz invariant manner in
$2+\epsilon$
dimensions and then regularized. As in [\Ell] we will not assume
Lorentz invariant integration [\Gates] so that some non Lorentz scalar momentum
integrals may be non vanishing. Otherwise we would find no contributions to the
$\beta$-functions at two loops. We will use the conventions of [\Pault] with
$c=0$ and define
$$
\epsilon^{\mu}_{\ \lambda}\epsilon^\lambda_{\ \nu}=-\delta^{\mu}_{\
\nu}+O(\epsilon^2)  \eqn\c
$$
in $D = 2+\epsilon$ dimensions. Other choices of regularization scheme are
equivalent to a field redefinition [\All].

Furthermore, in the vielbein frame the spin connection terms appearing in the
covariant derivatives, ${\omega^{(+)IJ}_=  = \omega^{(+)IJ}_k\dmm\Phi^k}$ and
${\omega^{(-)IJ}_+ = \omega^{(-)IJ}_k D_+\Phi^k}$, transform as $SO(d)$ gauge
potentials over $\Sigma^{(1,0)}$. This is  possible only in two spacetime
dimensions
where we may assign different connections to the left and right moving modes in
a
Lorentz and gauge invariant manner. The only gauge and Lorentz invariant
expressions
of the connection are the square of the field strength tensor and covariant
derivatives terms. The field strength tensor has mass dimension $3\over2$ and
so its
square can not be the coefficient of any divergence in (1,0) superspace
[\Alvarezo].
This also assumes that we use the convention \c\  [\All]. Therefore we may
replace
covariant derivatives by flat space derivatives while performing the loop
integrations as the connection terms will not contribute any Lorentz and gauge
invariant divergences. There is an important caveat here. If a diagram has a
positive
superficial degree of divergence the connection terms may well contribute to
the
divergences. This problem occurs at the 2 loop level. In such cases the
connection
terms in $\nabla^{(-)}_+$ covariant derivatives must be pulled back to the
background
fields when performing the $D_+$ algebra. This simplification greatly reduces
the
number of graphs which need to be considered.

As is well known there are potential sigma model anomalies arising from the
chiral structure
of (p,0) models which spoil the gauge and Lorentz invariance of the effective
action
[\Hullo,\Hullt,\Howeo]. This causes the connection terms to appear in
non invariant forms in the one loop effective action. It is, however, possible
to remove the anomalies to all orders in perturbation theory in a manner
consistent with (1,0) supersymmetry, by modifying the definition of $H_{ijk}$
through the addition of the Chern-Simons terms
$$\eqalign{
H_{ijk}\longrightarrow
&H_{ijk}+{3\a\over4}\Omega_3(A)-{3\a\over4}\Omega_3(\omega^{(-)}) \ , \cr
\Omega_3(A)_{ijk}&=A^{AB}_{[i}\partial_{j}A^{AB}_{k]}
+{2\over3}A^{AB}_{[i}A^{BC}_jA^{CB}_{k]}\ , \cr
\Omega_3(\omega^{(-)})_{ijk}&=\omega^{(-)IJ}_{[i}\partial_{j}\omega^{(-)IJ}_{k]}
+{2\over3}\omega^{(-)IJ}_{[i}\omega^{(-)JK}_j\omega^{(-)KI}_{k]}\ , \cr
}\eqn\Hmod
$$
where $\omega^{(-)}$ is the spin connection associated with
positive torsion [\Hullt,\Howeo]
$$
\omega^{(-)JK}_i=\omega^{JK}_i+H_{ijk}e^{jJ}e^{kK} \ .
$$
In this case we no longer have $dH=0$ but rather
$$
dH={3\over4}\a(F\wedge F-R^{(+)}\wedge R^{(+)}) \ .
\eqn\dh
$$
Thus the identity {\iden} is no longer valid, and furthermore, the connection
terms in the covariant derivatives will contribute Lorentz and gauge variant
divergences.
However, the effect of the anomalies on the two loop $\beta$-function has been
discussed
elsewhere for the massless (1,0) sigma model [\Callan, \Foakes]. In [\Callan]
it was
conjectured that the effect of the anomalies on the higher loop
$\beta$-functions was simply
to replace $H_{ijk}$ by the modification {\Hmod} and this was shown to
hold at the two loop level [\Ross,\Hamada]. The additional mass term considered
in
this paper, when expanded to arbitrary orders in the quantum fields, does not
contain any derivative operators. Hence the anomaly structure of the massive
(1,0)
supersymmetric sigma model is unchanged from the massless model and thus we may
use the analysis of [\Foakes]. Therefore for our purposes we may ignore the
contributions of the connection coefficients in \dh\  and continue to use
{\iden}.
After the computation is performed the effect of the one loop anomalies can be
accounted for by making the replacement {\Hmod} in the two loop
$\beta$-functions.

While the presence of the potential term in {\action} appears to act as a mass
term for
the fields, it in fact only does so, in the general (p,q)
supersymmetric case, for the $\phi^i$ and $\eta^i_+$ fields. We therefore treat
the potential term in  {\action} as a pure interaction term in the perturbative
expansion. In order to regulate the infrared divergences that occur we add a
mass $M$ into the propagator when  evaluating the divergent integrals. The
ultra-violet divergences can then be isolated  from the infrared ones. This
modification is a purely formal device achieved by simply using a  massive
propagator, in place of a massless one, in the momentum integrals. The infrared
divergent terms can then be ignored for the purpose of calculating the
$\beta$-functions. The massive propagators used are
$$\eqalign{
<0|T\{\xi^I(x,\theta^+_1)\xi^J(y,\theta^+_2)\}|0>&=
\delta^{IJ}\Dp(x,\thetap_1)\delta(\thetap_1-\thetap_2)\Delta(x-y)\, \cr
<0|T\{\zeta^A(x_1,\theta^+_1)\zeta^B(x_2,\theta^+_2)\}|0>
&=i\delta^{AB}\dminus^x\Dp(x_1,\thetap_1)\delta(\thetap_1-\thetap_2)\Delta(x_1-x_2)\ ,
\cr}
$$
where $\Delta(x_1-x_2)$ is the Feynman
propagator for a free bosonic field of mass $M$
$$
(\square+M^2)\Delta(x_1-x_2)=\delta(x_1-x_2)\ .
$$

The possible divergent one loop graphs are shown in figure 1. Note that triple
lines
represent background fields, solid lines $\xi^i$  propagators and dashed lines
$\zeta_-^a$ propagators. A slash on a propagator represents the insertion of a
momentum factor. However, only figure 1a yields a divergent integral after the
$D_+$
algebra is performed. Thus the $1\over\epsilon$ one loop divergences come
only from the tadpole graphs and are
$$\eqalign{ \Gamma^{(1,1)}_{Div}={-iI}\int\! d^2xd\thetap\{
-R^{(+)}_{ij}\Dp\Phi^i\dminus\Phi^j
&+{1\over2}i\Psima\hat{\nabla}^{(+)k}F^{ab}_{ki}\Dp\Phi^i\Psimb   \cr
&+{1\over2}im\hat{\nabla}_i\hat{\nabla}^is_a\Psima\}\ .     }\eqn\divone
$$
The divergent tadpole integral $I$ is given in the appendix.

Thus the requirements for one loop finiteness
are
$$
R^{(+)}_{ij}=0\, \ \ \ \ \ \
\hat{\nabla}^{(+)k}F^{ab}_{ki}=0\ ,\ \ \ \ \ \
\hat{\nabla}^2s_a=0\ ,
\eqn\finite
$$
The first two equations in {\finite} are
the well known one loop finiteness equations for the  heterotic sigma model,
which
may be rewritten as the Einstein and Yang-Mills field equations with the
antisymmetric field strength as a source [\Callan]. The last equation is the
new
contribution due to the potential $s_a$, not discussed in other supersymmetric
calculations. The presence of similar massive terms has, however, been
discussed in
the bosonic case in connection to the tachyon of bosonic string theory
[\Paffuti,\Osborn].

In order to renormalize the theory we define the renormalized fields
$g^r_{ij}$,
$b^r_{ij}$,  $A^{ra}_{i{\ }b}$ and $s^r_a$ in terms of the bare ones
$g^0_{ij}$,
$b^0_{ij}$, $A^{0a}_{i{\ }b}$ and $s^0_a$ with the divergences from each level
of loop
diagrams  subtracted off. In $2+\epsilon$ dimensions $g^0_{ij}$, $b^0_{ij}$ and
$A^{0a}_{i{\ }b}$ all have mass dimension $\epsilon$ while $ms^0_a$ has
dimension
$1+\epsilon$. Hence we write [\Alvarezo]
$$\eqalign{
g^0_{ij}&=\mu^\epsilon\left( g^r_{ij}-\sum_{\nu=1}^\infty
{\epsilon^{-\nu}g^\nu_{ij}}
\right)\ , \cr
b^0_{ij}&=\mu^\epsilon\left( b^r_{ij}-\sum_{\nu=1}^\infty
{\epsilon^{-\nu}b^\nu_{ij}}  \right)\ , \cr
A^{0a}_{i{\ }b}&=\mu^\epsilon\left( A^{ra}_{i{\ }b}
-\sum_{\nu=1}^\infty {\epsilon^{-\nu}A^{\nu a}_{i{\ }b}}
\right)\ , \cr
s^0_a&=\mu^{1+\epsilon}\left( s^r_a-\sum_{\nu=1}^\infty
{\epsilon^{-\nu}s^\nu_a}
\right)\ , \cr
}\eqn\poles
$$
where $g^\nu_{ij}$, $b^\nu_{ij}$, $A^{\nu a}_{i{\ }b}$ and $s^\nu_a$ are the
$1\over\epsilon^\nu$ divergent contributions to $g_{ij}$, $b_{ij}$, $A^a_{i{\
}b}$ and
$s_a$ respectively, calculated from all levels of loop diagrams. In \poles\ we
have
included the mass with $s_a$ since we
wish to include the classical contribution to the conformal anomaly. By
demanding that the bare fields do not depend on the (arbitrary) renormalization
scale $\mu$, it is well  known that the $\beta$-functions can be derived from
the
$\nu=1$ terms only in  {\poles} [\Alvarezo]. The higher order poles being
calculable
from the $1\over\epsilon$  terms via t'Hooft's pole equations.

Following this procedure the one loop divergences in {\divone} give rise to the
$\beta$-functions
$$\eqalign{
\beta^{(g)}_{ij}&=\mu{dg^r_{ij}\over{d\mu}}={\a}(R_{ij}-H_{imn}H_j^{\
mn})+O(\a^2)\ ,
\cr
\beta^{(b)}_{ij}&=\mu{db^r_{ij}\over{d\mu}}=-{\a}\nabla^kH_{kij}+O(\a^2)\ ,
\cr
\beta^{(A)ab}_{i}&=\mu{dA^{rab}_{i}\over{d\mu}}=-{\a\over2}(\hat{\nabla}^kF^{ab}_{ki}
+H^{\ kj}_{i}F^{ab}_{kj})+O(\a^2)\ ,\cr
\beta^{(s)}_{a}&=\mu{ds^r_a\over{d\mu}}=s_a-{\a\over2}\hat{\nabla}^2s_a+O(\a^2)
\
, \cr }
\eqn\bfuncone
$$
The factors of $\a$ have been inserted to enable us keep track of the
different loop contributions.

\subsection{(2,0) Supersymmetry}

In [\Hullth] and [\GP] the conditions for the model to admit an additional
off-shell
and on-shell left handed supersymmetry respectively, were found. The
requirements in
the case of zero potential [\Hullth,\Howet] are that ${\cal M}$ is a complex
manifold
with complex structure $I$, $g$ Hermitian with respect to $I$. In addition, the
holonomy of the connection $\Gamma^{(+)}$ must be a subgroup of $U(d/2)$.
Off-shell
supersymmetry further requires that $(\Xi,\hat{I},h)$ is also an Hermitian
manifold. Off-shell (2,0) supersymmetry of the mass terms requires
that, in the complex coordinates ($\mu=1,...,d/2$), ($\alpha,...,n/2$)
associated
with the complex structures ($I,{\hat I}$), the potential satisfies [\Hullth]
$$
s^\alpha={1\over2}\hat{I}_{r{\ }\beta}^\alpha(M^\beta_r-L^\beta_r)\ ,
\eqn\sML
$$
where $r=1,2$, $\hat{\nabla}\hat{I}_{r{\ }\beta}^\alpha=0$,
$\hat{\nabla}_{\overline{\mu}} L^\alpha_r=0$ and  $\hat{\nabla}_\mu
M^\alpha_r=0$. From
these conditions it clearly follows that
$$\eqalign{
\hat{\nabla}^2M^\alpha_r&=g^{\overline{\mu}\mu}
\hat{\nabla}_{\overline{\mu}}\hat{\nabla}_\mu M^\alpha_r=0 \ ,\cr
\hat{\nabla}^2 L^\alpha_r&=g^{\mu\overline{\mu}}
\hat{\nabla}_\mu\hat{\nabla}_{\overline{\mu}} L^\alpha_r=0 \ ,\cr}
$$
and hence
$$
\hat{\nabla}^2s^\alpha=0\ .
$$
Thus one sees that the mass term is one loop finite in the case of off-shell
(2,0) supersymmetry. In the case of on-shell supersymmetry, however, there are
many fewer restrictions on the target bundle [\GP]. In particular $\Xi$
need not be complex or even dimensional. Thus we are unable to conclude that
the mass
terms are in general one loop finite.

We will show by power counting arguments in section 6 below, that for off-shell
(2,0) models the mass term contributes no logarithmic divergences at any order.
The
$\beta$-function $\beta^{(s)}_a=s_a$ is then exact to all orders of
perturbation
theory.

\subsection{(1,1) Supersymmetry}

We now turn our attention to the case of (1,1) supersymmetry. Here the anomaly
{\Hmod} vanishes and, as will be shown in section 5, the $\beta$-functions
{\bfuncone}
receive no two loop corrections. For the model to admit a right handed
supersymmetry
[\Hullth], we must identify  $\,\Xi\,$ with the tangent bundle $T{\cal M}$ and
${\hat \nabla}_i$ with $\nabla^{(-)}_i$. We introduce a vielbein frame
$E_a{}^i$, with inverse $E^a_{\ i}$, such that
$$
h_{ab}=E_a{}^i E_b{}^j g_{ij} \ , \ \
\ \ \  {\nabla}^{(-)}_j E_a{}^i = 0 \ .  \eqn\vb
$$
The curvature and field strength are now simply related through
$$
F^{ab}_{ij} = E^{am}E^{bn}R^{(+)}_{ijmn}  \ .
\eqn\oneone
$$
With this identification the condition $R^{(+)}_{ij}=0$ along with the Bianchi
identity \Bianchi\ implies
$$
{\hat \nabla}^{(+)k}F^{ab}_{ki} = 0 \ ,
$$
so we need only ensure $R^{(+)}_{ij}=0$ for the massless sectors. In this case
$S_{\Phi}$ and $S_{\Psi}$ can be combined using the (1,1) superfield
$$\eqalign{
{\cX}^i &= \phi^i + \theta^+ \eta^i_+ + \theta^- \psi^i_- +
\theta^- \theta^+ F^i \cr
& = \Phi^i + \theta^- \Psi^i_- \cr }
\eqn\onesupfield
$$
where $\psi^i_- = E_a^{\ i}\psi^a_-$, $F^i = E_a^{\ i}F^a$ and
$\Psi^i_- = E_a^{\ i}\Psi^a_-$. $\eta^i_+$ and $\psi_-^i$ can now be thought of
as
the left and right handed components of a single spinor. In \onesupfield\
$\theta^-$ is another anticommuting spinorial coordinate similar to $\theta^+$
with
the corresponding superderivative  $$
D_- = {\partial\over\partial\theta^-} + i\theta^-\dmm \ ,
\eqn\Dmin
$$
which anticommutes with $D_+$.
The first two terms of the action \action\ can then be written as
$$
S_{\Phi+\Psi} = \int\ d^2x d\theta^+d\theta^-
(g_{ij}({\cX})+b_{ij}({\cX}))D_+{\cX}^i
D_-{\cX}^j \ . \eqn\onesupact
$$
The mass term $S_m$, however, can not always be written with (1,1) superfields.

The potentials consistent with (1,1)
supersymmetry are given by [\Hullth]
$$
s_a=E_a{}^i s_i \ , \ \ \ \ \
s_i=u_i-X_i\ ,
\eqn\sai
$$
where $X_i$ is a (possibly vanishing) Killing vector of ${\cal M}$;
$\nabla_{(i}
X_{j)}=0$ and $u_i$ is a one form satisfying
$$
X^kH_{ijk} =  \nabla_{[i} u_{j]}\ ,
\eqn\udef
$$
and the restriction [\Hullth,\Papa]
$$
X^iu_i=0\ .
\eqn\rest
$$
In the general case $X^i \ne 0$. The (1,1) supersymmetry algebra then contains
a
central charge and the massive terms in \action\ cannot be written in (1,1)
superfield form [\Hullth].

As can be seen from the definition of $u_i$ {\udef}, there are an infinite
number
of $u_i$ for a given $X_i$ in the
definition of $s_i$ which is reminiscent of the gauge freedom in
Electromagnetism with {\rest} as a gauge fixing condition. However, different
choices
of the function $u_i$ will lead to physically distinct potentials $s_i$. From
{\vb},
{\sai} and {\udef} one can derive $$\eqalign{
{\hat \nabla}^2 s_a&=E_a^{\ i}\nabla^{(-)2}s_i \cr
&=E_a{}^k(\nabla_k\nabla^i u_i + R^{(+)}_{jk}u^j + R^{(+)}_{jk}X^j)\cr
&=E_a{}^k\nabla_k\nabla^i u_i \ ,\cr}
\eqn\use
$$
when $R^{(+)}_{ij}=0$.

For a given $s_i$ we can find  another solution $u'_i$ to
{\udef} so the theory defined by $s'_i = u'_i - X_i$ is one loop finite. We
simply
take $u_i$ to be $u'_i= u_i +\nabla_i \lambda$ where, for an
arbitrary constant $\sigma$,
$$
\nabla^2 \lambda=\sigma-\nabla^i u_i \ .
$$
It follows that
$$
{\nabla^{(-)}}^2 s'_i=0 \ .
$$
However, we must satisfy the restriction {\rest} as well.
This can be done by requiring $\lambda$ to
satisfy the boundary condition $X^i\nabla_i\lambda=-X^iu_i$.

To see that this is possible, we can choose coordinates $(t,x^m)$ such that
$X={\partial\over{\partial t}}$ then $\lambda$ satisfies
$\nabla^2\lambda=\sigma-\nabla^i u_i$ and ${\partial\lambda\over{\partial
t}}=-u_t$.
This is analogous to choosing the Coulomb gauge in Electromagnetism.
Furthermore this
uniquely determines the function $u_i$, provided $X^i \ne 0$.

If $\sigma\ne 0$ then we may
rescale $m\rightarrow \sigma m$ and $s_i \rightarrow \sigma^{-1}s_i$ to absorb
$\sigma$ into $m$ and $s_i$. Thus for $X^i \ne 0$ there exist only two
physically distinct
potentials $s_a^{(\sigma)}$ which are one loop finite, provided
$R^{(+)}_{ij}=0$. If we assume $\cal M$ to  be
asymptotically flat and $s_a$ to vanish at infinity, we are forced to take
$\sigma=0$ and
we are left with a unique choice for $u_i$. Since
$X_i$ is Killing, $\nabla^i X_i=0$. Thus it follows from {\use} that the (1,1)
supersymmetric sigma model with potential $s_a$ is finite at one loop if and
only if
$\nabla^i s_i=\sigma$ and $R^{(+)}_{ij}=0$.

Given any target space ${\cal M}$ we can
choose $X=0$ so that {\rest} is trivially satisfied. Only in this case may we
write
$S_m$ in (1,1) superfield form as

$$
S_m = \int\ d^2x d\theta^+d\theta^- m\lambda({\cX}) \ .
$$
Then, $s_i$ is given by
$$
s_i=u_i = \nabla_i\lambda\
\eqn\Soneone
$$
and is finite at one loop precisely when
$$
\nabla^2\lambda=\sigma \ .
\eqn\zeroX
$$
Solutions to {\zeroX}, with $\sigma=0$, can be found by choosing $\lambda$ as
the real
part of a holomorphic function. In this case there are clearly an infinite
number of
physically distinct potentials which are one loop finite. The classical vacua
of the theory
are given by the critical points of $\lambda$. This is consistent with the
general
requirements of the $N=2$ nonrenormalization theorems (in the special case that
the model
possess (2,2) supersymmetry), where $\lambda$ is interpreted as the
superpotential in the
F-term, thus providing a check on our calculations.

Lastly, it we briefly outline what happens in models with (2,2) supersymmetry.
Since these models are special cases of (1,1) sigma models, we know that {\use}
holds.
However, as there also exists a second left handed supersymmetry we know that
$\nabla^2 s_a=0$ automatically. Thus for (2,2) supersymmetry we must have
$\nabla^iu_i=0$. Since $u_i$ is holomorphic [\Hullt] this is easily seen to be
the
case, because in complex coordinates $\nabla_{\bar{\mu}}u_\mu=0$ hence
$\nabla^\mu
u_\mu = g^{\bar{\mu} \mu}\nabla_{\bar{\mu}} u_\mu = 0$.

\subsection{The One Loop effective Potential}

To complete our discussion of the one loop quantization of the
massive supersymmetric sigma model we now calculate the effective potential.
For this it is sufficient to set $\eta^i_+=\psi^a_-=0$ and
fix $\phi^i$ to be constant. While the sigma model is most readily defined by
the
superspace action {\action}, in this simple case we choose to calculate the
effective
potential using ordinary component fields rather than superfields. To this end
we
expand the action {\potential} without integrating over the auxiliary field
$F^a$ (i.e.
with $V = -ms_aF^a - F^aF^a$), to second order in the quantum component fields.
This yields,
after integrating out the quantum auxiliary fields,
$$\eqalign{
S^{(2)} = \int\! d^2 x & \left\{ \delta_{IJ}\dpp A^I \dmm A^J +
i\delta_{IJ}\omega_+^I\dmm\omega_+^J - i \delta_{AB}\chi^A_-\dpp\chi^B_-
\right. \cr
&\left.
+ m{\hat \nabla}_Is_A\omega^I_+\chi^A_-
- {1\over4}(m^2{\hat \nabla}_I s_A {\hat \nabla}_J s^A -
2mF^A{\hat \nabla}_{(I}{\hat \nabla}_{J)}s_A) A^I A^J \right\} \ ,\cr}
\eqn\bfqa
$$
where we have referred all fields to the vielbein frames and $A^I$,
$\omega^I_+$
and $\chi^A_-$ are the quantum fields for $\phi^I$, $\eta^I_+$ and $\psi^A_-$
respectively.

It is useful to define the matrices
$$\eqalign{
M_{IA} &= m{\hat \nabla}_I s_A \ , \cr
M^2_{IJ} &= m^2 {\hat \nabla}_I s_A {\hat \nabla}_J s^A \ ,\cr
K^F_{IJ} &= {1\over2}(m^2{\hat \nabla}_I s_A {\hat \nabla}_J s^A
-2mF^A{\hat \nabla}_{(I}{\hat \nabla}_{J)}s_A)\ . \cr}
\eqn\bfqb
$$
$M_{IA}$ and $K_{IJ}$ appear in the action {\bfqb} as mass matrices for the
(left
handed) fermions and bosons respectively. Only in the case of (1,1)
supersymmetry do
they provide a mass for the right handed fermions.

{}From {\bfqa} the propagators for the component quantum fields can be read
off. We do
not add a mass term for these fields as we did above since in general
the infrared divergences cancel in the effective potential - although we will
discuss
an exception to this below. In
\bfqa\  there are boson-boson and fermion-fermion interactions. The effective
potential can be found by summing over all one loop diagrams with zero external
momentum. There are two types of graph to consider, the purely bosonic loop
2a and the purely fermionic loop 2b. It is a straight forward
calculation to determine their contributions to be (in Euclidean space)
$$\eqalign{
5a) &= {\a\over4\pi}\int\ {d^2p} \
{\rm Tr\ ln}\left( \delta_{IJ} + {K^F_{IJ}\over p^2}\right) \cr
5b) &= -{\a\over4\pi}\int\ {d^2p} \
{\rm Tr\ ln}\left( \delta_{IJ} + {M^2_{IJ}\over 2p^2}\right)
\ . \cr}
\eqn\bfqc
$$
where the trace is over the manifold indices.
We must also include the counter term graphs in \bfqc . However these are just
the
graphs in figure 5 with only one and two vertices respectively. The momentum
integral
in \bfqc\ can then be performed and we arrive at the one loop correction to the
potential
$$\eqalign{
V_{eff} &= -ms_AF^A - h_{AB}F^AF^B \cr
&+ {\a\over4}{\rm Tr}\left[
K^F_{IJ} - K^F_{IJ}{\rm ln}\left({K^F_{IJ}\over \mu^2}\right)
- {1\over2}M^2_{IJ}
+ {1\over2}M^2_{IJ}{\rm ln}\left({M^2_{IJ}\over 2\mu^2}\right) \right] \ .}
\eqn\bfqd
$$

We now find the equation of motion of $F^A$ to be
$$
F^A = -{1\over2}ms^A + {m\a\over8}{\rm Tr} \left[
{\hat \nabla}_{(I}{\hat \nabla}_{J)}s^A{\rm ln}{K_{IJ}\over\mu^2} \right]
+ {\cal O}(\hbar^2) \ ,
\eqn\bfqe
$$
where
$$
K_{IJ} = {m^2\over2}({\hat \nabla}_I s_A {\hat \nabla}_J s^A
+s^A{\hat \nabla}_{(I}{\hat \nabla}_{J)}s_A) \ .
\eqn\bfqf
$$
Substituting this into \bfqd\ yields
$$\eqalign{
V_{eff} &= {1\over4}m^2s_As^A+ {\a\over4}{\rm Tr} \left[
K_{IJ} - K_{IJ}{\rm ln}{K_{IJ}\over\mu^2} - {1\over2}M_{IJ}^2
+ {1\over2}M_{IJ}^2{\rm ln}{M_{IJ}^2\over2\mu^2}
\right]
\cr
& + {m^2\a^2\over64}
\left( {\rm Tr} \left[ {\hat \nabla}_{(I}{\hat \nabla}_{J)}s^A
{\rm ln}{K_{IJ}\over\mu^2} \right] \right)^2 \ .\cr }
\eqn\bfqg
$$
We have kept the last term in \bfqg\ since, although it is of order $\a^2$, it
is determined by one loop corrections and is needed to ensure that the
effective
potential is positive.

Suppose that ${\hat \nabla}_Is^A$ is invertible. We may then expand \bfqg\ near
the classical vacuum $\phi_{cl}$ as, $y^I = {O}(\a)$,
$$\eqalign{
V_{eff}(\phi_{cl}+y) &= {m^2\over4} \left(
{\hat \nabla}_Is^A(\phi_{cl})y^I
-{\a\over4} {\rm Tr}\left[
{\hat \nabla}_{(I}{\hat \nabla}_{J)}s^A(\phi_{cl})
{\rm ln}{M^2_{IJ}(\phi_{cl})\over2\mu^2} \right]
\right)^2  \cr
&+ \a^2V_2(\phi_{cl}) + {\cal O}(\a^3) \ , \cr}
\eqn\bfqh
$$
where $V_2$ represents the higher loop contributions to $V_{eff}$. Since
${\hat \nabla}_Is^A$ is invertible, we can solve for $y^I$ so that the first
term in
\bfqh\ vanishes and $V_{eff}$ is minimized. The
classical vacuum is then shifted by $y^I$, of order $\a$ and
the vacuum energy is $\a^2V_2(\phi_{cl}) + {\cal O}(\a^3)$. Since supersymmetry
must be preserved in this case (see below) we conclude that $V_2(\phi_{cl})=0$,
justifying the inclusion of the $O(\a^2)$ term in \bfqg .

If ${\hat \nabla}_Is^A$ is degenerate at $\phi_{cl}$, so that the fermion mass
matrix $M^2_{IJ}$ has zero modes then the effective potential diverges
logarithmically there. It is tempting to view the presence of massless fermions
as an
indication that the theory dynamically breaks supersymmetry, with the massless
fermions interpreted as Goldstone modes. It is sometimes incorrectly stated
that
supersymmetry can not be broken perturbatively due to the nonrenormalization
theorems. In fact what actually prevents perturbative corrections to the vacuum
energy is a non vanishing Witten index, which counts the  number of bosonic
minus the number of fermionic zero energy states. A non vanishing  index
therefore implies the existence of a supersymmetric (zero energy) vacuum state.
As is
well known the Witten index is  a topological invariant, equal to the Euler
number of
$\cal M$ in the (1,1) supersymmetry case, and therefore cannot receive any
quantum corrections, including non perturbative effects. Vacuum states where
$M^2_{IJ}$ has a zero mode, however, do not contribute to the Witten index
(Euler number) and so may in principle be removed by quantum effects. This
issue
has been raised and discussed some time ago where it was concluded that no
such breaking of supersymmetry occurred (see [\Mur] and the references
therein).
However, the models discussed there claimed to break supersymmetry even in
cases
where the Witten index was nonzero. What we are describing here actually
corresponds to the case $a=0$ of [\Mur] which, to the best of
our knowledge, has not been discussed.

Unfortunately it is  precisely the existence of the massless fermions which
causes
the logarithmic infrared divergence in the effective potential. This renders
the
loop expansion invalid near the classical vacuum. Thus, despite the fact that a
straightforward analysis shows the effective potential to be non vanishing at
the
vacuum, we cannot conclude that supersymmetry has been perturbatively broken.
If
we add a mass term in the propagator as an infrared regulator, as we have done
above
when calculating the $\beta$-functions, then we find that the effective
potential
is well behaved and the vacuum energy is indeed lifted above zero. However,
this is
not surprising as the addition of the mass term explicitly breaks
supersymmetry. In
order to substantiate the claim that supersymmetry is perturbatively broken we
must
find a reliable approximation in which to work. Although in a weaker sense
supersymmetry is broken at 1 loop as standard perturbation theory becomes
untenable about such a vacuum state. In the non-perturbative regime, dynamical
breaking of supersymmetry can be seen using the $1/N$ expansion in the large
$N$ limit
[\Sal] for similar types of two dimensional fields theories.

\chapter{Two Loop Renormalization}

In this section we proceed to calculate the two loop contributions to the gauge
and
Lorentz invariant parts of the $\beta$-functions, bearing in mind the
discussion in
the  previous section regarding the effect of the one loop anomalies. In order
to
calculate the $\beta$-functions to two loop order we must first expand the
action
{\action} to  fourth order in the quantum fields. This calculation is greatly
simplified by using the  algorithm we developed in section 3. To third and
fourth
order, the action is  $$\eqalign{
S^{(3)}_\Phi={-i\over3}\int\! d^2xd\thetap
&\left\{\tilde{\nabla}^{(+)}_{(m}R^{(+)}_{|i|kl)j}D_+\Phi^i\dminus\Phi^j\xi^k\xi^l\xi^m
+2R^{(+)}_{i(kl)j}\delpp\xi^i\dminus\Phi^j\xi^k\xi^l  \right. \cr
& \left.
+ 2R^{(+)}_{i(kl)j}D_+\Phi^i\delmm\xi^j\xi^k\xi^l
+ 2H_{ijk}\delpp\xi^i\delmm\xi^j\xi^k
\right\} \ ,}
$$
$$\eqalign{
S^{(3)}_\Psi={-i\over6}\int\! d^2xd\thetap &\left\{
i\Psima\Psimb(\tilde{\nabla}^{(-)}_{(l}\tilde{\nabla}^{(+)}_kF^{ab}_{|i|j)}
+R^{(+)p}_{i(kl}F^{ab}_{|p|j)})D_+\Phi^i\xi^j\xi^k\xi^l \right. \cr
& \left.
- 2i\Psima\Psimb\hat{\nabla}_{(k}F^{ab}_{j)i}\delpp\xi^i\xi^j\xi^k
-6i\Psima F^{ab}_{ij}\delpp\xi^i\xi^j\zeta_-^b \right. \cr
& \left.
- 6i\Psima\hat{\nabla}^{(+)}_{(k}F^{ab}_{|i|j)}D_+\Phi^i\xi^j\xi^k\zeta_-^b
+ 6iF^{ab}_{ij}\Dp\Phi^i\xi^j\zeta_-^a\zeta_-^b
\right\} \ ,}
$$
$$
S^{(3)}_m={-i\over6}\int\! d^2xd\thetap \left\{
im\hat{\nabla}_{(i}\hat{\nabla}_j\hat{\nabla}_{k)}s_a\Psima\xi^i\xi^j\xi^k
+3im\hat{\nabla}_{(i}\hat{\nabla}_{j)}s_a\zeta_-^a\xi^i\xi^j
\right\} \ ,
$$
$$\eqalign{
S^{(4)}_\Phi={-i\over12}\int\! d^2xd\thetap
&\left\{
(4R^{(+)\ p}_{i(mn}R^{(+)}_{|p|kl)j}+
\tilde{\nabla}^{(-)}_{(n}\tilde{\nabla}^{(+)}_mR^{(+)}_{|i|kl)j})
\Dp\Phi^i\dminus\Phi^j\xi^k\xi^l\xi^m\xi^n
\right. \cr
& \left.
+ (6\nabla_{(l}H_{k)ij}+4R_{i(kl)j})\delpp\xi^i\delmm\xi^j\xi^k\xi^l
\right. \cr
& \left.
+ 3\nabla_{(m}^{(+)}R^{(+)}_{|i|kl)j}\Dp\Phi^i\delmm\xi^j\xi^k\xi^l\xi^m
\right. \cr
& \left.
+ 3\nabla_{(m}^{(-)}R^{(-)}_{|i|kl)j}\dminus\Phi^i\delpp\xi^j\xi^k\xi^l\xi^m
\right\} \ ,}
$$
$$\eqalign{
S^{(4)}_\Psi={-i\over24}\int\! d^2xd\thetap &\left\{
12i\tilde{\nabla}^{(+)}_{(k}F^{ab}_{|i|j)}\Dp\Phi^i\xi^j\xi^k\zeta_-^a\zeta_-^b
\right. \cr
& \left.
-8i\Psima(\tilde{\nabla}^{(-)}_{(l}\tilde{\nabla}^{(+)}_kF^{ab}_{|i|j)}
+R^{(+)p}_{i(kl}F^{ab}_{|p|j)}
)\Dp\Phi^i\dminus\Phi^j\xi^j\xi^k\xi^l\zeta_-^b\right.\cr
& \left.
+16i\Psima\hat{\nabla}_{(k}F^{ab}_{j)i}\delpp\xi^i\xi^j\xi^k\zeta_-^b
+12i F^{ab}_{ij}\delpp\xi^i\xi^j\zeta_-^a\zeta_-^b \right. \cr
& \left.
+i\Psima\Psimb(3\tilde{\nabla}^{(+)}_{(k}F^{ab}_{|p|j}R^{(+)\ p}_{|i|lm)}
+\tilde{\nabla}^{(+)}_{(m}\tilde{\nabla}^{(-)}_l\tilde{\nabla}^{(+)}_kF^{ab}_{|i|j)}
\right. \cr
& \left.
+\tilde{\nabla}^{(+)}_{(m}R^{(+)p}_{|i|kl}F^{ab}_{|p|j)}-4H_{p(k}^{\ \
n}F^{ab}_{|n|j}R^{(+)\ p}_{|i|lm)})\Dp\Phi^i\xi^j\xi^k\xi^l\xi^m
\right. \cr
& \left.
+i\Psima\Psimb(2\hat{\nabla}_{(l}\hat{\nabla}_kF^{ab}_{|i|j)}
-2H_{i(l}^{\ \ p} \hat{\nabla}_{k} F^{ab}_{|p|j)}
+\tilde{\nabla}^{(-)}_{(l}\tilde{\nabla}^{(+)}_kF^{ab}_{|i|j)}
\right. \cr
& \left.
+F^{ab}_{p(j}R^{(+)p}_{|i|kl)})\delpp\xi^i\xi^j\xi^k\xi^l
\right\} \ ,}
$$
$$
S^{(4)}_m={-i\over24}\int\! d^2xd\thetap \left\{
im\hat{\nabla}_{(i}\hat{\nabla}_j\hat{\nabla}_k\hat{\nabla}_{l)}
s_a\Psima\xi^i\xi^j\xi^k\xi^l
+4im\hat{\nabla}_{(i}\hat{\nabla}_j\hat{\nabla}_{k)}s_a\zeta_-^a\xi^i\xi^j\xi^k
\right\} \ .
$$
In the above expansions we have introduced a "twisted" covariant derivative
${\tilde \nabla}^{(\pm)}$ defined on target manifold bundles as
$$
\tilde{\nabla}^{(+)}_k T^{a_1...a_m}_{i_1...i_n}=\hat{\nabla}_k
T^{a_1...a_m}_{i_1...i_n} -  H^{\ \ \ p}_{ki_1}T^{a_1...a_m}_{pi_2...i_n}
+ H^{\ \ \ p}_{ki_2}T_{i_1pi_3...i_n}\dots +
(-1)^{n}H^{\ \ \ p}_{ki_n}T^{a_1...a_m}_{i_1...p}\ ,
\eqn\twder
$$
and similarly for $\tilde{\nabla}^{(+)}$ with $H_{ijk}$ replaced by
$-H_{ijk}$.

The counter terms, derived from {\divone} and expanded to second order in the
quantum
fields are of the form
$$\eqalign{
S^{(2)}_D={-i\over2\pi\epsilon} \int\! d^2xd\thetap &\left\{
C^2_{-(ij)}\xi^i\xi^j +
C^2_{=ij}\delpp\xi^i\xi^j+C^2_{+jk}\delmm\xi^j\xi^k
\right. \cr & \left.
+C^2_{ij}\delpp\xi^i\delmm\xi^j+ D^2_{+ab}\zeta_-^a\zeta^b_- +
D^2_{ai}\zeta^a_-\xi^i
\right. \cr & \left.
+D^2_{-ai}\zeta^a_-\delpp\xi^i+ D^2_{{\ne}ai}\zeta^a_-\delmm\xi^i \right\} \ ,}
\eqn\countertwo
$$
However, as will be seen below we only need to know the $C^2_{ij}$,
$C^2_{+ij}$, $C^2_{=ij}$, $D_{\ne ai}$ and $D^2_{- ai}$ coefficients in order
to
calculate the contributions of {\countertwo} to the two loop $\beta$-functions.
These
coefficients are
$$\eqalign{
C^2_{ij}&=-R^{(+)}_{ij}\ , \cr
C^2_{+jk}&=-(\nabla_k R^{(+)}_{ij}+H_{ik}^{\ \ m}R^{(+)}_{mj})\Dp\Phi^i \ ,\cr
C^2_{=jk}&=-(\nabla_k R^{(-)}_{ij}-H_{ik}^{\ \ m}R^{(-)}_{mj})\dminus\Phi^i\
,\cr
D^2_{-ai}&=i\Psima\hat{\nabla}^{(+)k}F^{ab}_{ki} \ , \cr
D^2_{\ne a i} &=0 \ . \cr}
$$

Simple power counting shows that the only divergent two loop diagrams have at
most 3 vertices. By dimensional analysis, it can been seen that no vertices
coming from the  expansion of the mass term $S_m$ in {\action} can contribute
to
the renormalization of  the $g_{ij}, b_{ij}$ and $A^a_{i\ b}$ fields. This is
because any such diagram would  necessarily involve terms with
$m\Dp\Phi^i\dminus\Phi^j$ or $m\Psima\Psimb\Dp\Phi^i$  respectively. However,
both of these terms have mass dimension $5/2$ and hence the  corresponding
graphs must have a negative superficial degree of divergence. Thus any
divergences must come from divergent subgraphs, but these are canceled,
according to  Hepp's theorem, by the counter term graphs.

\subsection{$S_\Phi$ Renormalization}

Since no vertices coming from the expansion of $S_m$ and only one graph from
$S_\Psi$
contribute to the $g_{ij}$ and $b_{ij}$ renormalization, we have only the
graphs in
figure 3 to calculate. These are
$$\eqalign{
3a)&={i}IJ \int\! d^2xd\thetap
\left\{R^{kl}R^{(+)}_{i(kl)j}\Dp\Phi^i\dminus\Phi^j \right\} \ ,\cr
3b)&=-{4i\over9}K \int\! d^2xd\thetap
\left\{H^{k}_{\ mn}H^{lmn}R^{(+)}_{i(kl)j}\Dp\Phi^i\dminus\Phi^j \right\} \
,\cr
3c)&=-{2i\over9}\left(1 - {\epsilon\over2}\right)L \int\! d^2xd\thetap
\left\{(R^{(+)(kl)m}_{i}R^{(+)}_{j(kl)m}\right.
\cr
&\left.\ \ \ \ \ \ \ \ \ \ \ \ \ \ \ \ \ \ \ \ \ \ \ \ \ \ \ \ \ \ \ \ \ \ \ \
\ \ \ \ \ \
-R^{(+)(kl)m}_{i}R^{(+)}_{j(lm)k})\Dp\Phi^i\dminus\Phi^j \right\}  \ ,\cr
3d)&={i\over4}\left(1 - {\epsilon\over2}\right)L \int\! d^2xd\thetap
\left\{F^{ab}_{ik}F^{ab\ k}_{\ \ j}\Dp\Phi^i\dminus\Phi^j \right\} \ ,\cr
3e)&=-{i\over9}\left(1 - {\epsilon\over2}\right)L \int\! d^2xd\thetap
\left\{\nabla^{(-)}_iH_{klm}\nabla^{(-)}_jH^{klm}\Dp\Phi^i\dminus\Phi^j
\right\}  \ ,\cr
3z)&={i\over{2\pi\epsilon}}J \int\! d^2xd\thetap
\left\{R^{(+)(kl)}R^{(+)}_{i(kl)j}\Dp\Phi^i\dminus\Phi^j \right\} \ ,\cr
}
$$
where the divergent integrals $I$, $J$, $K$ and $L$ are given in the appendix.
Only one of the three possible combinations of curvature terms appears in graph
3c,
the others are either finite or lead to $1\over\epsilon^2$ contributions only.
The $D_+$ algebra, however, reduces the momentum integral to one that is not a
Lorentz scalar. To evaluate this integral we contract the internal momentum
with the external $\dpp\Phi^i$ field and express everything in terms of
$\eta_{\mu\nu}$ and $\epsilon_{\mu\nu}$. The integration then yields the extra
factor
of $\epsilon$ appearing at the front of 3c, 3d and 3e and also the $\dmm\Phi^i$
field. In this way we obtain non vanishing $1\over\epsilon$ poles. If we had
assumed
symmetric Lorentz integration then these graphs would vanish.

It is not hard to see, using {\Revenodd}, that the $1\over\epsilon$ pole terms
from
graphs (3a,3b,3z) cancel each other. We now find, after some
algebraic manipulations, that the two loop
$1\over\epsilon$ pole divergences of (3c,3d,3e) are
$$
\Gamma^{(1,2)}_{Div\ \Phi}=
{-i\over{64\pi^2\epsilon}} \int\! d^2xd\thetap
\left\{R^{(+)}_{iklm}R^{(+)klm}_{j} - F^{ab}_{ki}F^{abk}_{\ \ \ \ j}
\right\} \Dp\Phi^i\dminus\Phi^j\ .
\eqn\sphitwo
$$
We may now proceed, as in section 4, to calculate the two loop
$\beta$-functions
$\beta^{(g)}_{ij}$ and $\beta^{(b)}_{ij}$. They are easily seen to be
$$\eqalign{
\beta^{(g)}_{ij}&={\a}R^{(+)}_{(ij)}
+{\a^2\over8}\left( R^{(+)}_{iklm}R^{(+)klm}_{j}
-F^{ab}_{ki}F^{abk}_{\ \ \ \ j}
 \right) + O(\a^3)\ ,\cr
\beta^{(b)}_{ij}&={\a}R^{(+)}_{[ij]} +O(\a^3)\ .\cr }
\eqn\betatwogb
$$

\subsection{$S_\Psi$ Renormalization}

The vertices from $S_\Phi$ and $S_\Psi$ both contribute to the renormalization
of
$S_\Psi$. The non vanishing graphs are given in figure 4. Following a similar
analysis to the $S_{\Phi}$ renormalization we obtain the contributions
$$\eqalign{
4a)&={2i\over9}K \int\! d^2xd\thetap
\left\{ i\Psima H^k_{\ mn}H^{jmn}\hat{\nabla}^{(+)}_k F^{ab}_{ij}
\Dp\Phi^i\Psimb
\right\} \ ,\cr
4b)&=-{2i\over9}K \int\! d^2xd\thetap\left\{ i\Psima H_j^{\ mn}
R^{(+)}_{ikmn}F^{abkj}\Dp\Phi^i\Psimb\right\} \ ,\cr
4c)&=-{i\over2}K' \int\! d^2xd\thetap
\left\{i\Psima H^{kmn}F^{ac}_{ik}F^{cb}_{mn}\Dp\Phi^i\Psimb \right\} \ ,\cr
4d)&={-i}IJ \int\! d^2xd\thetap
\left\{i\Psima F^{acj}_i \hat{\nabla}^kF^{cb}_{jk} \Dp\Phi^i\Psimb \right\} \
,\cr
4e)&={3i\over8}IJ \int\! d^2xd\thetap
\left\{i\Psima  g^{mn}\nabla^{(+)}_{(k}R^{(+)}_{|i|mn)j}F^{abkj}\Dp\Phi^i\Psimb
\right\} \ ,\cr
4f)&=-{i\over2}IJ \int\! d^2xd\thetap \left\{i\Psima R^{kj}
\hat{\nabla}^{(+)}_kF^{ab}_{ij}\Dp\Phi^i\Psimb \right\} \ ,\cr
4g)&=-{i\over8}IJ \int\! d^2x d\thetap \left\{ i\Psima
\nabla_lH^{lkj}{\hat \nabla}^{(-)}_iF^{ab}_{kj}
D_+\Phi^i \Psimb \right\}\ ,\cr
4w)&=-{i\over4\pi\epsilon}J \int\! d^2xd\thetap \left\{ i\Psima
R^{(+)(jk)}\hat{\nabla}^{(+)}_k F^{ac}_{ij}\Dp\Phi^i\Psimb \right\} \ ,\cr
4x)&=-{i\over4\pi\epsilon}J \int\! d^2xd\thetap
\left\{i\Psima(\nabla_k R^{(+)}_{ij}+H_{ik}^{\ \
m}R^{(+)}_{mj})F^{abkj}\Dp\Phi^i\Psimb \right\} \ ,\cr
4y)&=-{i\over2\pi\epsilon}J\int\! d^2xd\thetap
\left\{i\Psima\hat{\nabla}^{(+)k}F^{ac}_{kj}F^{cbj}_i\Dp\Phi^i\Psimb \right\}\
, \cr
4z)&={i\over16\pi\epsilon}J \int\! d^2x d\thetap \left\{ i\Psima (
R^{(+)kj}{\hat \nabla}^{(-)}_iF^{ab}_{kj}+\nabla^{(-)}_iR^{(+)kj}F^{ab}_{kj}
)D_+\Phi^i \Psimb \right\}\ . \cr}
$$
{}From these graphs, the integrals given in the appendix and the identities
{\Revenodd}, {\syms} and {\iden} we find that the $1\over\epsilon$ poles of
graphs
(4a,4f,4w) and (4c,4d,4y) have completely canceled with each other.
Furthermore after some tedious algebra, the $1\over\epsilon$ poles from the
graphs
(4b,4e,4x) and the second term in 4z also cancel, while the first
term in 4z cancels with 4g. Thus we find
$$
\Gamma^{(1,2)}_{Div\ \Psi}=0
$$
and arrive at the Yang-Mills two loop $\beta$-function
$$
\beta^{(A)ab}_i=-{\a\over2}\hat{\nabla}^{(+)k}F^{ab}_{ki}+O(\a^3) \ .
\eqn\betatwoA
$$

\subsection{$S_m$ Renormalization}

Finally we consider the renormalization of the mass term $S_m$. The only
diagrams
contributing to the mass renormalization at two loops are given in figure 5.
They
can be calculated to be
$$\eqalign{
5a)&={i\over2}IJ \int\! d^2xd\thetap
\left\{imR_{ij}\hat{\nabla}^i\hat{\nabla}^js_a\Psima\right\} \ ,\cr
5b)&=-{2i\over9}K\int\! d^2xd\thetap
\left\{imH_{ikl}H_j^{\ kl}\hat{\nabla}^i\hat{\nabla}^js_a\Psima\right\} \ ,\cr
5c)&=-{i\over4}K' \int\! d^2xd\thetap
\left\{imH_i^{\ jk}F^{ab}_{jk}\hat{\nabla}^is_a\Psima\right\} \ ,\cr
5d)&={i\over2}IJ \int\! d^2xd\thetap
\left\{im\hat{\nabla}^kF^{ab}_{ki}\hat{\nabla}^is_a\Psima\right\} \ ,\cr
5y)&={i\over4\pi\epsilon}J \int\! d^2xd\thetap
\left\{imR^{(+)}_{(ij)}\hat{\nabla}^i\hat{\nabla}^js_a\Psima\right\} \ , \cr
5z)&={i\over4\pi\epsilon}J \int\! d^2xd\thetap
\left\{im\hat{\nabla}^{(+)k}F^{ab}_{ki}\hat{\nabla}^is_b\Psima\right\} \ .\cr}
$$
These divergences can now be added up, using expressions for the divergent
integrals
in the appendix and {\Revenodd} to give the total two loop contribution to
the $1\over\epsilon$ pole divergence. One finds that the $1\over\epsilon$ terms
in
(5a,5b,5y) and (5c,5d,5z) completely cancel leaving only $1\over\epsilon^2$
poles.
Hence
$$
\Gamma^{(1,2)}_{Div\ m}=0
$$
and so we find the $\beta$-function $\beta^{(s)}_a$,
calculated to two loop order, to be
$$
\beta^{(s)}_a=s_a-{\a\over2}\hat{\nabla}^2s_a + O(\a^3)\ .
$$

\subsection{Field Redefinitions}

There is an inherent quantum mechanical ambiguity in the above calculation
caused by
the possible introduction of $O(\a)$ finite local counter terms, equivalent to
a
change in renormalization scheme. These terms have no effect at the one loop
level,
however the one loop diagrams constructed from them will alter the two loop
$\beta$-functions. We will therefore end our discussion by
considering the effect that the addition of such terms to the action \action\
has
on the $\beta$-functions found above. As discussed above the addition of finite
local
counter terms is  need for the effect of the sigma model anomaly to be included
in
them case of chiral (p,0) supersymmetry. In addition they are also need to
preserve (4,0) supersymmetry in perturbation theory and ensure that the
off-shell
(4,0) supersymmetric sigma models are ultraviolet finite [\Howthree].

The addition
of finite local counter terms is tantamount to making a redefinition of
$g_{ij}$,
$b_{ij}$, $A^a_{i\ b}$ and $s_a$ that appear in {\action} to
$$\eqalign{
g_{ij} & \longrightarrow g_{ij} + \a{\bg}_{ij} \
,\cr
 b_{ij} &
\longrightarrow b_{ij} + \a{\bb}_{ij} \ ,\cr
 A^a_{i\ b} & \longrightarrow A^a_{i\ b}
+ \a{\bA}^a_{i\ b} \ ,\cr
s_a & \longrightarrow s_a + \a{\bs}_a \ .\cr}
\eqn\redefs
$$
As the potential terms do not effect the other beta functions,  it is of little
interest here to consider $s_a$ redefinitions. Lets us suppose then  that only
$g_{ij}$, $b_{ij}$ and $A^{a}_{i\ b}$ have been redefined as in  {\redefs}. The
connection $\Gamma^{(-)}$ is shifted to
$$
\Gamma^{(-)i}_{\ \ \ jk} \longrightarrow \Gamma^{(-)i}_{\ \
\ jk} + \a  {\bG}^i_{\ jk} + O(\a^2)\ ,
$$
where
$$
{\bG}_{ijk} = {1\over2}(\nabla_j {\bg}_{ik}
+\nabla_k {\bg}_{ij}-\nabla_i {\bg}_{jk}) - {3\over2}\nabla_{[k}{\bb}_{ij]}\ ,
$$
and we use the original metric $g_{ij}$ to raise and lower indices. The Yang-
Mills field strength $F^{ab}_{ij}$ is shifted to
$$
F^{ab}_{ij} \longrightarrow F^{ab}_{ij}
+ {\a}{\hat \nabla}_{[i}{\bA}^{ab}_{j]} + O(\a^2) \ .
$$
A straightforward
calculation shows that the two loop $\beta$-functions become (to
$O(\a^2)$)
$$\eqalign{
{\beta'}^{(g)}_{ij} &= \beta^{(g)}_{ij} + \a^2({\tilde
\nabla}^{(-)k}{\bG}_{k(ij)} - \nabla^{(-)}_{(i|}\nabla^k{\bg}_{k|j)}) +
\a^2\nabla^{(-)}_{(i}v_{j)}  \ ,\cr
{\beta'}^{(b)}_{ij} &= \beta^{(b)}_{ij} + \a^2({\tilde
\nabla}^{(-)k}{\bG}_{k[ij]} - \nabla^{(-)}_{[i|}\nabla^k{\bg}_{k|j]}) +
\a^2\nabla^{(-)}_{[i}v_{j]}   \ ,\cr
{\beta'}^{(A)ab}_{i} &= \beta^{(A)ab}_{i}
+ {\a^2\over2}\bG_{kij}F^{abjk}
- {\a^2\over2}({\bA}^{ka}_{\ \ c}F^{cb}_{ij} + {\bA}^{kb}_{\ \ c}F^{ac}_{ij}
+ {\hat \nabla}^k{\hat \nabla}_{[k}{\bA}^{ab}_{i]}) \cr
&\ \ \ \ \ \ \ \ \ \ \  + {\a^2\over2}v^kF^{ab}_{ki}\ ,\cr
{\beta'}^{(s)}_a &=\beta^{(s)}_a + {\a^2\over2}({\hat \nabla}^k{\bA}_k^{ab}
+ {\bar A}^{\ \ \ b}_{ka}{\hat \nabla}^ks_b)
+ {\a^2\over2}v^k{\hat \nabla}_k s_a\ , \cr}
$$
where $v_k = g^{ij}{\bG}_{kij}$ and ${\tilde \nabla}^{(-)}$ is defined in
\twder.

We are also free to redefine the background fields $\Phi^i$ and $\Psi^a_-$.
Indeed if
we consider the diffeomorphism generated by the vector $v^i$, $\Phi^i
\rightarrow
\Phi^i + \a^2v^i$ accompanied by a gauge transformation with parameter
$u^{a}_{\ b} =
-\a^2v^i A^{a}_{i\ b}$ then the $\beta$-functions are (to $O(\a^2)$)
$$\eqalign{
{\beta''}^{(g)}_{ij} & = \beta^{(g)}_{ij} + \a^2({\tilde
\nabla}^{(-)k}{\bG}_{k(ij)} - \nabla^{(-)}_{(i|}\nabla^k{\bg}_{k|j)})\ ,\cr
{\beta''}^{(b)}_{ij} & = \beta^{(b)}_{ij} + \a^2({\tilde
\nabla}^{(-)k}{\bG}_{k[ij]} - \nabla^{(-)}_{[i|}\nabla^k{\bg}_{k|j]})  \ ,\cr
{\beta''}^{(A)ab}_{i} &= \beta^{(A)ab}_{i}
+ {\a^2\over2}\bG_{kij}F^{abjk}
-{\a^2\over2}({\bA}^{ka}_{\ \ c}F^{cb}_{ij} + {\bA}^{kb}_{\ \ c}F^{ac}_{ij}
+ {\hat \nabla}^k{\hat \nabla}_{[k}{\bA}^{ab}_{i]})
\ ,\cr
{\beta''}^{(s)}_a &=\beta^{(s)}_a + {\a^2\over2}({\hat \nabla}^k{\bA}_k^{ab}
+ {\bar A}^{\ \ \ b}_{ka}{\hat \nabla}^ks_b) \ . \cr}
\eqn\betaredef
$$
It is by the above procedure that the effect of the sigma model anomalies can
be
included by setting
$$
{\bG}_{ijk} = {3\over4}\Omega_{3ijk}(\omega^{(-)})-{3\over4}\Omega_{3ijk}(A) \
,
$$
where $\Omega_3$ is defined in \Hmod .

\chapter{Concluding Remarks}

In this paper we calculated the two loop $\beta$-functions of the general
massive
(p,q) supersymmetric sigma model using (1,0) superfields. At two loops the
$\beta$-functions are (without taking account of any potential sigma model
anomalies)
$$\eqalign{
\beta^{(g)}_{ij}&={\a}(R_{ij}-H_{ikl}H_j^{\ kl})
+{\a^2\over8}\left( R^{(+)}_{iklm}R^{(+)klm}_{j}
-F^{ab}_{ki}F^{abk}_{\ \ \ j}
 \right)\ ,\cr
\beta^{(b)}_{ij}&=-{\a}\nabla^kH_{kij} \ , \cr
\beta^{(A)ab}_i&=-{\a\over2}(\hat{\nabla}^{k}F^{ab}_{ki}+H^{\
kj}_{i}F^{ab}_{kj}) \ , \cr
\beta^{(s)}_a&=s_a-{\a\over2}\hat{\nabla}^2s_a  \ . \cr}
\eqn\betafinal
$$
These results are in agreement
with previous calculations in special cases where various background
fields vanish [\Foake,\Foakes,\Ell].

As is well known [\Alvarezo,\All,\Jones,\VdV,\GM], when we restrict the model
so that
it possess (1,1) supersymmetry, the two loop divergences  from $R^{(+)}_{ijkl}$
and
$F^{ab}_{ij}$ in {\betafinal} cancel. This can be seen by setting
$A_i^{ab}=\omega^{(+)ab}_i$, so that the anomaly {\Hmod} vanishes (in fact it
can be
arranged, by the addition of finite local counter terms, that any spin
connection
differing from $\omega^{(+)}$ by covariant term may be used [\Hullf]).
Furthermore, we have $F^{ab}_{ij}=E^{am}E^{bn}R^{(+)}_{ijmn}$ and can identify
($\eta^i_+,E_a^{\ i}\psi^a_-$) as the left and right handed components of a
single
spinor. It then clearly follows that the $\a^2$ terms in {\betafinal} vanish.
There
will in general be higher loop divergences [\GVZ].

\subsection{Ultra-violet Behaviour At All Orders}

Here we would like here to discuss ultra-violet behaviour to all orders for the
general off shell (p,q) supersymmetric massive sigma model. As is well known
theories
with $N=4$ supersymmetry are often finite as this places very strict conditions
on
the possible counter terms. Following the power counting argument of [\Howet],
we
now show that off-shell (4,q) supersymmetric massive sigma models are
perturbatively
ultra-violet finite to all orders, for $q\leq4 $. We will also see that
there are no mass renormalizations to all orders of perturbation theory for the
off-shell (2,q) models.

For an off shell (p,q) supersymmetric sigma model in two dimensions the
superspace
measure is
$$
d^2xd^p\theta^+d^q\theta^-\ .
$$
This has Lorentz weight $p-q$ and mass dimension ${1\over2}(p+q)-2$. Thus any
logarithmically divergent counter term must have mass dimension
$2-{1\over2}(p+q)$ and
Lorentz weight $q-p$. The possible divergences
are of the form
$$
\Gamma_{Div} \sim m^\alpha{\cal
O}(\partial,D_-,D_+)F(\Phi)(\Psi_+)^r(\Psi_-)^s\ ,
$$
where $\cal O$ is a differential operator and F a scalar function. F may
contain
derivatives of $\Phi$, so long as they are in in scalar combinations. Hence we
may
choose F in such a way that it contains the maximum number of derivatives. In
this case
$\cal O$ is either of the form ${\cal O} \sim (D_+)^a$ or  ${\cal O} \sim
(D_-)^b$.

Now the counter term $\Gamma_{Div}$ has Lorentz weight
$$
[\Gamma_{Div}]_l=q-p=[{\cal O}]_l + r - s\ ,
$$
and mass dimension,
$$
[\Gamma_{Div}]_m=2-{1\over2}(p+q)=\alpha + [{\cal O}]_m + [F]_m +
{1\over2}(r+s) \ .
$$
However, all of {\cal O}'s Lorentz weight must come from worldsheet
derivatives, and our choice of F implies that
$$
[{\cal O}]_m={1\over2}|[{\cal O}]_l| \ .
$$
Hence, the mass dimension of F must satisfy
$$\eqalign{
[F]_m&=2-\alpha-{1\over2}(p+q)-{1\over2}(r+s)-[{\cal O}]_m \cr &=
2-\alpha-{1\over2}(p+q)-{1\over2}(r+s)-{1\over2}|((p+r)-(q+s))| \cr &\leq
2-\alpha-r-p\
, \cr} \eqn\bound
$$
where we have used the inequality $|a-b|\ge|a|-|b|$. As $\Gamma_{Div}$ has a
non
negative degree of divergence and there are no negative mass dimensional
constants, $[F]_m\ge0$. Thus, for $p>2$ there are no possible logarithmically
divergent  counter terms. As $p=3$ implies $p=4$, we see that off-shell (4,q)
supersymmetric sigma models are perturbatively ultra-violet finite to all
orders.

A quick look at \betafinal\ appears to show a contradiction with
this claim. There is a two loop contribution to the metric $\beta$-function
which does not in general vanish in the case of chiral (4,0) supersymmetry.
This
problem has been recognized before and is resolved by the observation that
(4,0)
supersymmetry is broken in perturbation theory [\Howthree]. To remedy this
requires the addition of finite local terms (ie. field redefinitions) at each
order of perturbation theory to restore (4,0) supersymmetry. The metric and
antisymmetric tensor fields of the finite theory are then not those that appear
in {\action} but differ by terms of higher order in $\a$. In other words in an
appropriate regularization scheme the $(4,0)$ models are
ultra-violet finite.

The inequality {\bound} has a further application to the case of massive (2,q)
models. Here it follows that $\alpha=0$ for any counter term and therefore
there
are no  possible $S_m$ counter terms. Thus there are no mass renormalizations
to all
orders of  perturbation theory for the off-shell (2,q) models.

\subsection{Conformal Invariance and String Theory}

Finally we would like to make some comments concerning the relationship of
massive
supersymmetric sigma model to string theory. As is easily seen from the action
{\action} the presence of the mass term $ms_a\Psima$ breaks conformal
invariance at
the classical level and this appears as the first term in $\beta^{(s)}_{a}$.
{}From the
point of view of string theory the sigma model $\beta$-functions become
equations of
motion for the (bosonic) spacetime fields $g_{ij},b_{ij},A^{a}_{i\ b}$ and
$\varphi$
[\Callan]. Here $\varphi$ is the  dilaton which is absent from our calculations
as it
does not couple to a flat worldsheet and hence does not appear action \action .
If we
likewise consider $s_a$ as a spacetime field the minus sign in
$\beta^{(s)}_{a}$
implies that it is a tachyon with (mass)$^2 \sim -1/\a$. Although, since a
tachyon
does not appear in the spectrum of the superstring, it's interpretation as a
spacetime field is problematic. Usually in order to render the theory spacetime
supersymmetric the GSO projection is performed which then removes the tachyon
from
the spectrum of the superstring. However, one may wish to consider string
theories
with worldsheet supersymmetry but no spacetime supersymmetry, in which case the
tachyon would be in the physical spectrum.

The vanishing of the $\beta$-functions \bfuncone\  only ensures that
rigid scale invariance is preserved. As is discussed in
[\Paffuti,\Tseytlin,\Osborn]
full conformal invariance of the sigma model occurs when the complete
$\beta$-functions, including any additional contributions from the dilaton,
vanish.
The central charge is then given by the dilaton $\beta$-function, which is a
constant
by virtue of the Curci-Paffuti relation [\Paffuti,\Tseytlin,\Osborn].
Furthermore the
Curci-Paffuti relation enables the dilaton $\beta$-function, and it's
contributions
to the other $\beta$-functions, to be calculated by flat worldsheet techniques
[\Osborn,\Jack]. Only then can the complete conditions for conformal invariance
found
and the central charge be made to vanish. Thus, even though the action
{\action} is
not classically conformally invariant it may be possible that the quantum
theory is.

Any statement regarding a nontrivial solution of $\beta^{(s)}_a=0$ (or it's
generalization to include the dilaton) must be made carefully as we are
comparing terms
of different order in $\a$ and therefore we cannot assume that the loop
expansion is
meaningful. A special case is offered by massive linear sigma models for which
$F^{ab}_{ij}=H_{ijk}=R_{ijkl}=\varphi=0$. $\beta^{(s)}_a$ is then the only non
vanishing $\beta$-function. Since the superspace measure
$d^2xd\theta^+$ has mass dimension $-{3\over 2}$ and each vertex carries a
factor of
$m$, the only divergent contributions to the effective action come from graphs
with a
single vertex. Of these however, only the one loop tadpole graph has a
${1\over\epsilon}$ pole which contributes to the $\beta$-function. Thus
$\beta^{(s)}_a$ in \bfuncone\ is exact at order $\a$ to all orders of
perturbation
theory and receives no other (perturbative) contributions.\foot{It would be
interesting to know if this were still true had we only assumed
$F^{ab}_{ij}=R^{(+)}_{ijkl}=\varphi=0$. Here an extension of the arguments in
[\Muhkit] based on the algorithm \rules\ again show that only $\beta^{(s)}_a$
is
nonzero.}

The author would like to thank H. Osborn, G. Papadopoulos, P.K. Townsend and M.
Wes
for their advice and Trinity
College Cambridge for financial support.

\endpage

\chapter{Appendix}

In order to regulate the divergent integrals we use the conventions of
[\Gates,\Pault]. The
divergent integrals used above, after taking the limit $M^2\rightarrow 0$ and
ignoring any infrared divergences, are found by dimensional regularization to
be
$$\eqalign{
I&=\int\! {d^Dk\over{i(2\pi)^D}}{1\over{-k^2+M^2}}\cr
&=-{1\over{2\pi\epsilon}}+O(\epsilon) \ , \cr
J&=\int\!
{d^Dk\over{i(2\pi)^D}}{k_{\neq}k_=\over{[-k^2+M^2]}}{1\over{[-(p-k)^2+M^2]}}
\cr
&={1\over{4\pi\epsilon}}+{1\over{4\pi}} + O(\epsilon) \ , \cr
K&=-K_1+K_2-K_3+K_4-K_5+K_6 \cr
&={1\over{16\pi^2\epsilon^2}}-{9\over{32\pi^2\epsilon}}
+O(\epsilon^0)\ , \cr
K'&=K_3-K_6 \ ,\cr
&={1\over{8\pi^2\epsilon^2}}+{1\over{4\pi^2\epsilon}}+O(\epsilon^0)\ ,\cr
L&=\int\! {d^Dq\over{i(2\pi)^D}} \int\! {d^Dk\over{i(2\pi)^D}}
{k_{\neq} k_=\over{[-k^2+M^2]}}{1\over{[-q^2+M^2]}}{1\over{[-(p-k-q)^2+M^2]}}
\cr
&=-{1\over{8\pi^2\epsilon^2}} + O(\epsilon^0) \ ,\cr}
$$
where
$$\eqalign{
K_1&=\int\! {d^Dq\over{i(2\pi)^D}} \int\! {d^Dk\over{i(2\pi)^D}}
\left\{{k_{\neq} \over{[-k^2+M^2]}}{q_=\over{[-q^2+M^2]}} \right. \cr
&\left. \ \ \ \ \ \ \ \ \ \ \ \ \ \ \ \ \ \ \ \ \ \ \ \ \ \ \ \ \ \ \ \
{(p-k-q)_=\over{[-(p-k-q)^2+M^2]}}{(p'+k)_{\neq}\over{[-(p'+k)^2+M^2]}}
\right\} \cr
&={1\over{32\pi^2\epsilon}}+O(\epsilon^0) \ ,\cr
K_2&=\int\! {d^Dq\over{i(2\pi)^D}} \int\! {d^Dk\over{i(2\pi)^D}}
\left\{{k_{\neq} \over{[-k^2+M^2]}}{q_= q_=\over{[-q^2+M^2]}} \right. \cr
&\left. \ \ \ \ \ \ \ \ \ \ \ \ \ \ \ \ \ \ \ \ \ \ \ \ \ \ \ \ \ \ \ \
{1\over{[-(p-k-q)^2+M^2]}}{(p'+k)_{\neq}\over{[-(p'+k)^2+M^2]}}
\right\} \cr
&={1\over{16\pi^2\epsilon^2}}+{1\over{32\pi^2\epsilon}}+ O(\epsilon^0)\ ,\cr
K_3&=\int\! {d^Dq\over{i(2\pi)^D}} \int\! {d^Dk\over{i(2\pi)^D}}
\left\{{k_{\neq} \over{[-k^2+M^2]}}{q_{\neq} q_=\over{[-q^2+M^2]}} \right. \cr
&\left. \ \ \ \ \ \ \ \ \ \ \ \ \ \ \ \ \ \ \ \ \ \ \ \ \ \ \ \ \ \ \ \
{1\over{[-(p-k-q)^2+M^2]}}{(p'+k)_=\over{[-(p'+k)^2+M^2]}}
\right\} \cr
&={1\over{16\pi^2\epsilon^2}}+{1\over{8\pi^2\epsilon}}+ O(\epsilon^0)\ ,\cr
K_4&=\int\! {d^Dq\over{i(2\pi)^D}} \int\! {d^Dk\over{i(2\pi)^D}}
\left\{{k_{\neq} \over{[-k^2+M^2]}}{q_{\neq} q_=\over{[-q^2+M^2]}} \right. \cr
&\left. \ \ \ \ \ \ \ \ \ \ \ \ \ \ \ \ \ \ \ \ \ \ \ \ \ \ \ \ \ \ \ \
{(p-k-q)_=\over{[-(p-k-q)^2+M^2]}}{1\over{[-(p'+k)^2+M^2]}}
\right\} \cr
&=O(\epsilon^0)\ ,\cr
K_5&=\int\! {d^Dq\over{i(2\pi)^D}} \int\! {d^Dk\over{i(2\pi)^D}}
\left\{{k_{\neq} \over{[-k^2+M^2]}}{q_= q_=\over{[-q^2+M^2]}} \right. \cr
&\left. \ \ \ \ \ \ \ \ \ \ \ \ \ \ \ \ \ \ \ \ \ \ \ \ \ \ \ \ \ \ \ \
{(p-k-q)_{\neq}\over{[-(p-k-q)^2+M^2]}}{1\over{[-(p'+k)^2+M^2]}}
\right\} \cr
&={1\over{32\pi^2\epsilon}}+ O(\epsilon^0)\ ,\cr
K_6&=\int\! {d^Dq\over{i(2\pi)^D}} \int\! {d^Dk\over{i(2\pi)^D}}
\left\{{k_{\neq} \over{[-k^2+M^2]}}{q_=\over{[-q^2+M^2]}} \right. \cr
&\left. \ \ \ \ \ \ \ \ \ \ \ \ \ \ \ \ \ \ \ \ \ \ \ \ \ \ \ \ \ \ \ \
{(p-k-q)_{\neq}\over{[-(p-k-q)^2+M^2]}}{(p'+k)_=\over{[-(p'+k)^2+M^2]}}
\right\} \cr
&=-{1\over{16\pi^2\epsilon^2}}-{1\over{8\pi^2\epsilon}}+ O(\epsilon^0)\ .\cr}
$$

\endpage

\refout

\endpage

%\insert figure 1 here

\midinsert
\epsfysize= 3cm
\centerline{\epsffile{sigma1.eps}}
\centerline{\it Figure 1: contributions to $\Gamma^{(1,1)}_{Div}$}
\endinsert

%\insert figure 2 here

\midinsert
\epsfysize=4cm
\centerline{\epsffile{sigma5.eps}}
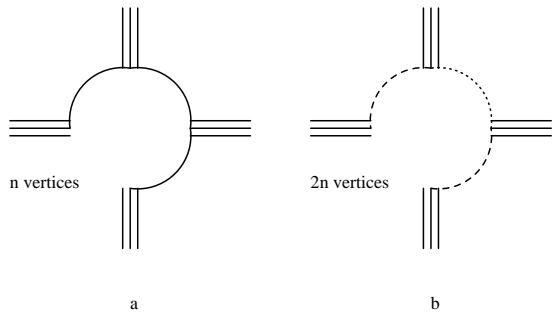
\centerline{\it Figure 2: 1 loop contributions to $V_{eff}$}
\endinsert

%\insert figure 3 here

\midinsert
\epsfysize= 8cm
\centerline{\epsffile{sigma2.eps}}
\centerline{\it Figure 3: contributions to $\Gamma^{(1,2)}_{Div\ \Phi}$}
\endinsert

%\insert figure 4 here

\midinsert
\epsfysize=16cm
\centerline{\epsffile{sigma3.eps}}
\centerline{\it Figure 4: contributions to $\Gamma^{(1,2)}_{Div\ \Psi}$}
\endinsert

%\insert figure 5 here

\midinsert
\epsfysize=8cm
\centerline{\epsffile{sigma4.eps}}
\centerline{\it Figure 5: contributions to $\Gamma^{(1,2)}_{Div\ m}$}
\endinsert

\end